\documentclass[12pt]{article}

\usepackage[utf8]{inputenc}
\usepackage[T1]{fontenc}
\usepackage{amsmath, amssymb, amsfonts}
\usepackage{graphicx}
\usepackage{caption}
\usepackage{authblk}
\usepackage{geometry}
\usepackage{hyperref}
\usepackage{booktabs}
\usepackage{mhchem}
\usepackage{orcidlink}
\usepackage{lmodern}
\usepackage{setspace}
\usepackage[numbers]{natbib}  
\usepackage{subcaption}

\title{Machine Learning Workflow for Analysis of High-Dimensional Order Parameter Space: \\ A Case Study of Polymer Crystallization from Molecular Dynamics Simulations}

\author[1]{Elyar Tourani\orcidlink{0009-0003-5181-6835}}
\author[1]{Brian J. Edwards\orcidlink{0000-0002-2378-5627} \thanks{Email: \texttt{bje@utk.edu}}}
\author[1]{Bamin Khomami\orcidlink{0000-0002-0091-2312} \thanks{Email: \texttt{bkhomami@utk.edu}}}
\affil[1]{Materials Research and Innovation Laboratory, Department of Chemical and Biomolecular Engineering, University of Tennessee, Knoxville, TN 37996, USA}

\date{\today}

\begin{document}

\maketitle

\newpage 

\begin{abstract}
Currently, identification of crystallization pathways in polymers is being carried out using molecular simulation-based data on a preset cut-off point on a single order parameter (OP) to define nucleated or crystallized regions. Aside from sensitivity to cutoff, each of these OPs introduces its own systematic biases. In this study, an integrated machine learning workflow is presented to accurately quantify crystallinity in polymeric systems using atomistic molecular dynamics (MD) data. Each atom is represented by a high-dimensional feature vector that combines geometric, thermodynamic-like, and symmetry-based descriptors. Low-dimensional embeddings are employed to expose latent structural fingerprints within atomic environments. Subsequently, unsupervised clustering on the embeddings is used to identify crystalline and amorphous atoms with high fidelity. After generating high-quality labels with multidimensional data, we use supervised learning techniques to identify a minimal set of order parameters that can fully capture this label. Various tests were conducted to reduce the feature set and it is shown that using only three order parameters, namely $q_6$, $\bar{S}_i$, and $p_2$, is sufficient to recreate the crystallization labels with great accuracy. Based on these observed OPs, the crystallinity index ($C$-index) is defined as the logistic regression model’s probability of crystallinity. This measure remains bimodal at all stages of the process and achieves $> 98\%$ classification performance (AUC). Notably, a model trained on one or a few snapshots enables efficient on-the-fly computation of crystallinity. Lastly, we demonstrate how the optimal $C$-index fit evolves during various stages of crystallization, supporting the hypothesis that entropy dominates early nucleation, while $q_6$ gains relevance in the later stages. This workflow yields a data-driven strategy for OP selection and provides a generalizable metric to monitor structural transformations in large-scale polymer simulations.
\end{abstract}


\newpage

\section{Introduction}
\label{sec:intro}

Studying the detailed crystallization pathways in polymer physics leads to improvements in materials and manufacturing processes \cite{VENABLES1994798, BOISTELLE198814,whitelam2015statistical}. 
However, the chain-like nature of polymers adds a layer of complexity to the crystallization process that is not present in small molecules. This complexity requires conformational and spatial order, which makes it challenging to fully understand the mechanisms of crystallite formation from polymer melts and solutions \cite{keller1952morphology, armitstead1992polymer, gedde1995polymer, sawyer2008polymer, luo2017molecular}.

Classical theories such as Lauritzen-Hoffman (LH) \cite{Hoffman1976} and Sadler-Gilmer (SD) \cite{sadler1984model}, along with concepts of multistage nucleation pathways \cite{olmsted1998spinodal,strobl2000melt,tang2017local}, have provided insight but still leave unresolved questions, including the precise role of precursor liquid structures, the dynamics of polymer segments entering and leaving crystallites, and the presence of orientational order prior to crystallization.

Recent nanoscale simulations of polymer crystallization have explored nucleation and growth dynamics
\cite{yamamoto2009computer, rutledge2013computer, 10.1063/1.3240202, yamamoto2001molecular, anwar2013crystallization, muthukumar2000modeling, 10.1063/1.4816707, yi2013molecular, anwar2015crystallization, nafar2020thermodynamically, hussain2024coarse},
often using angle-based OPs to identify local crystalline segments by evaluating segmental orientations \cite{esselink1994molecular, 10.1063/1.3240202, 10.1063/1.3608056, nicholson2016analysis, anwar2013crystallization},
which suffer from threshold sensitivity, have difficulties at chain ends or on the surface of a growing crystal cluster \cite{sadler1984model}.
Averaging techniques to address these issues can reduce precision, especially when characterizing cluster surfaces and merging nuclei.

Recently, localized thermodynamic-like descriptors, derived from atomic configurational entropy and enthalpy, have effectively distinguished crystalline phases in metals and polymers \cite{piaggi2017enhancing, piaggi2017entropy, nettleton1958expression, nafar2020thermodynamically}. Nafar Sefiddashti et al. \cite{nafar2020thermodynamically, nafar2020flow} studied flow-induced polymer crystallization under elongational flow conditions by noting a sharp change in the atomic configurational entropy.
They also demonstrated that this method's robust hyperparameter selection requires no prior knowledge of the system's physical state, such as chain conformational ordering angle.
However,  the method uses a modified approximate radial distribution function, $g(r)$ and its efficacy in identifying subtle structures, such as folds and bridges during quiescent crystallization, requires further validation.

Various other translational and orientational order parameters, such as Common Neighbor Analysis (CNA) \cite{faken1994systematic}  and its adaptive versions \cite{tsuzuki2007structural, stukowski2012structure}, Bond Angle Analysis (BAA) \cite{ackland2006applications}, Centrosymmetry Parameter (CNP) \cite{kelchner1998dislocation}, Voronoi tessellation \cite{finney1970random, tanemura1978geometrical}, and Steinhardt's Bond-Orientational Order (BOO) \cite{steinhardt1983bond}, have been widely applied in small molecules \cite{tanaka2019revealing} and occasionally polymers \cite{ORBi-21ffd385-366f-4535-a1db-8ac65d2246e9}.  Voronoi analysis and BOO parameters are among the few OPs that have been shown to help identify ordering within a polymeric system \cite{ORBi-21ffd385-366f-4535-a1db-8ac65d2246e9}.

Voronoi analysis adapts to particle volume changes during packing, but does not reveal details about the symmetry of a particle's neighbor list. Evidently, the difference in symmetrical features in an aligned and a folded section of the chain may hinder a more precise identification of the symmetry-related order parameters.

The BOO parameters
use spherical harmonics to identify local motifs, which were
used in many previous studies, as reviewed in \cite{musil2021physics}.
Despite their practicality, the conventional BOO parameters are greatly affected by the criteria chosen to define the neighbor list \cite{mickel2013shortcomings}.
The influence of per-particle values, which are not sufficiently sensitive to distinguish between geometrically similar Wyckoff sites in highly complex crystal structures, can obscure genuine structural changes arising from the system's underlying physics, underscoring the importance of thoughtfully evaluating methods for neighbor inclusion. These methods may include distance-based, number-based, or topology-based approaches, such as the solid angle-based nearest-neighbor algorithm (SANN) \cite{van2012parameter}, or Polyhedral Template Matching (PTM) \cite{larsen2016robust} based on the number of Voronoi cell faces. Although these analytical approaches show promise for small molecules \cite{musil2021physics}, their use in polymer crystallization remains unexplored.

Machine learning (ML) methods have shown significant potential for analyzing phases, transition states, and crystal growth pathways in a wide range of materials systems \cite{wang2018nonlinear, wang2016discovering, jadrich2018unsupervised, gasparotto2014recognizing, yang2018large,
helfrecht2020structure, boattini2019unsupervised, asoudeh2024pegylated, barakati2024reward, doi:10.1021/acs.chemrev.1c00021}.
Spellings and Glotzer \cite{spellings2018machine} developed
signatures inspired by BOO OPs, focusing on the principal axes of inertia, creating a high-dimensional feature space and using ML methods for the construction of phase diagrams of specific crystal structures. Martirossyan et al. \cite{doi:10.1021/acsnano.4c01290} further developed this using clustering of the PCA and Gaussian mixture model (GMM) to identify intricate local motifs and coordination complexities during crystallization. For polymeric systems, Bhardwaj et al. \cite{bhardwaj2024nucleation} recently employed self-supervised autoencoders to analyze local conformations and environmental fingerprints, uncovering latent structural patterns for polymers.
We believe that further refinement and benchmarking against conventional methods remain essential for developing more sensitive and robust crystallization OPs with a focus on their ability to be interpreted and generalized.

This study helps the community establish new ML-based workflows in which a high-dimensional feature space is turned into a more understandable signature. Our more unique contributions are,
(i) the use of the clustering boundaries to study the systematic biases of each OP regarding crystallization. Essentially, providing a simple pipeline to work back from the ML results to learn about features themselves.
(ii) Use of causal feature selection to keep all the information about the labels while reducing the computational burden,
(iii) Validation between stages (fitting the crystallinity index ($C$-index) at one time and testing it at other time points) to ascertain the generalization of the label construction plus feature selection process.
(iv) Show evidence that different order parameters explain different stages of polymer crystallization.
(v) By focusing on the ability to interpret, we establish a versatile framework that is readily adaptable to other measurement techniques or material systems.

Here, we focus on polymer crystallization in polyethylene (PE) using MD simulations, details of which are given in Section 2, along with methods to characterize the local environment.
Section 3 evaluates ML-driven descriptors against traditional parameters and provides guidelines for replicating multidimensional crystallization. Using this insight, the $C$-index is defined at the end of this section. The key findings and the broader implications are summarized in Section 4.

\section{Methods}
\label{sec:methodology}

\subsection{Molecular dynamics simulation details}\label{sec:methods_md}

The Siepmann-Karaboni-Smit (SKS) united atom (UA) potential \cite{siepmann1993simulating} is used to represent polyethylene chains. In this model, the terminal \ce{CH3} groups are modeled as head and tail beads, while the intermediate \ce{CH2} groups constitute the internal atoms of the chain. To address the challenges related to bonded interactions, the original rigid bond constraints are replaced with a harmonic potential, which improves the stability of the model and removes the need to apply bond constraints in the numerical integration scheme\cite{moore2000molecular, baig2006rheological, cui1996multiple, baig2005rheological, ionescu2006structure}. Although there are no definitive studies confirming the ability of these or similar UA force fields to accurately represent the lowest-energy microstructures in crystalline phases \cite{zhang2016review}, they remain valuable for investigating nucleation dynamics and crystal growth \cite{yamamoto2019molecular, nafar2020thermodynamically}. In this context, it is advisable to compare the modeled systems with manually optimized crystal structures, assessing their proximity to minimal-energy configurations. 

In the SKS model, non-bonded intermolecular and intramolecular interactions are quantified using the 6- to 12-Lennard-Jones (LJ) potential, defined as 
\begin{equation} 
U_{LJ}(r_{ij}) = 4 \epsilon_{ij} \left[\left(\frac{\sigma_{ij}}{r_{ij}}\right)^{12} - \left(\frac{\sigma_{ij}}{r_{ij}}\right)^6 \right]~, 
\label{energy_lj} 
\end{equation} 
where $\epsilon_{ij}$ represents the depth of the potential well for the interactions of particles $i$ and $j$. The values of $\epsilon / k_B$ are specified as $47$~K for \ce{CH2} groups and $114$~K for \ce{CH3} groups, with $k_B$ being the Boltzmann constant. The parameter $\sigma_{ij}$ denotes the distance at which the intermolecular potential is zero, standardized at 3.93~\AA~for both \ce{CH2} and \ce{CH3} groups. The variable $r_{ij}$ indicates the instantaneous distance between the particles $i$ and $j$. To consider this potential for two particles in the same chain, they should be separated by at least three other particles along the chain. A cutoff distance of $2.5 \, \sigma_{\rm {CH_2}}$ was implemented for all LJ potentials. The interaction parameters for disparate pairs of particles $i$ and $j$ were calculated using the Lorentz-Berthelot mixing rules, $\epsilon_{ij} = (\epsilon_i \epsilon_j)^{1/2}$ and $\sigma_{ij} = \frac{\sigma_i + \sigma_j}{2}$. A harmonic potential function provides the bond-stretching interaction energy: $U_{str}(l) = \frac{k_l}{2} (l - l_0)^2$, where $l_0$ is the equilibrium distance for the bond between two neighboring particles in a chain equal to $1.54 \ \AA$, and $l$ is the instantaneous distance between those two particles. The bond stretching constant $k_l$ is set to ($k_l / k_B=$) $452,900 \ \rm{K/ \AA^2}$. A harmonic potential function is also used to calculate the bond bending potential: $U_{bend}(\theta) = \frac{k_\theta}{2} (\theta - \theta_0)^2$, where $\theta$ is the angle created by three successive particles. The equilibrium angle is $\theta_0 = 114^o $ and the bond bending constant is $ k_{\theta} / k_B = 62,500 \ \rm{K/rad^2}$. The energy of the bond-torsion interaction was defined as $U_{tor}(\phi) = \sum_{m=0}^3 a_m (cos \phi)^m$. Here, $\phi$ is the angle between the $(i,j,k)$-plane and the $ (j,k,l)$-plane created by 4 successive bonded particles. The coefficients $ a_0 / k_B $ to $ a_3 / k_B $ were set to $ 1010$, $-2019$, $136.4$, and $3165$~K, respectively. 
Further details on the force field are available in the following references ~\cite{siepmann1993simulating, nafar2015individual, nafar2020thermodynamically}.

Molecular dynamics simulations were performed using LAMMPS \cite{plimpton1995fast, thompson2022lammps} under constant pressure and temperature conditions, maintaining pressure at one atmosphere with orthorhombic periodic boundary conditions, using the Nose-Hoover thermostat and barostat. For comparison purposes, we chose 60 n-pentacontahectane (C150) chains at $300$~K, achieving a $25\%$ supercooling level, aligned with previous studies \cite{yi2013molecular, anwar2013crystallization}. The system was equilibrated at $550$~K for $200$~ns to ensure thermal homogenization, then quenched to $300$~K to induce nucleation, where a single stable nucleation event was observed during cooling. The density measurements ranged from $0.76~{\rm{g/cm^3}}$ at $450$~K to $0.94~\rm{g/cm^3}$ through crystallization at $300$~K.
The analysis included 10 snapshots during the quenching phase to examine transitions in crystallinity using order parameters. 

We used an orthorhombic polyethylene lattice as a reference structure, 
that was constructed based on the crystallographic space group $Pnma\_D\_{2h}$, with lattice parameters set at $7.48$, $2.55$ (chain axis) and $4.97$~\AA, excluding hydrogen atoms. The carbon atoms were placed using Wyckoff positions of $0.033$ and $0.066$, resulting in a crystal density of $0.996~\rm{g/cm^3}$ \cite{gedde2019morphology, Pereira2008}.

\subsection{Definition of Order Parameters for Crystal Cluster Analysis}

To characterize the structural environments of polymer chains during crystallization, several geometric, orientational, and symmetry-based order parameters were utilized. In the following, we provide an overview of each descriptor used to create per-atom feature vectors for further analysis.

\subsubsection{Voronoi tessellation for local density calculation}

In a molecular dynamics simulation, Voronoi tessellation can offer a parameter-free geometric method to determine the neighbors of each particle in three dimensions. This technique quantifies the local density at the location of each particle $i$ by calculating the inverse of the volume of the Voronoi cell of the particle. The simulation box is divided into discrete regions, each uniquely associated with a single particle, encompassing all points closer to the particle than any other, and is defined as the Voronoi cell or volume of the particle. The number of faces of the Voronoi cell for each particle, which is the number of nearest neighbors of the central atom, can also be considered as a parameter(\texttt{nfaces}).

\subsubsection{Local angle-based crystallinity order parameter, \( p_2 \)}\label{sec:methods_p2}

The process of crystallization in polymers often involves the alignment of chain segments, a crucial step preceding the formation of crystalline structures. To quantitatively depict this alignment, the local \( p_2 \) order parameter is used for each particle \( i \), which is defined as 
\begin{equation}
p_2(i) = \left\langle \frac{3 \cos^2 \theta_{ij} - 1}{2} \right\rangle_j ~.
\label{eq:p2def}
\end{equation}
Here, \(\theta_{ij}\) represents the angle between the bond vector from bead \( i-1 \) to bead \( i+1 \) and the corresponding vector from bead \( j-1 \) to bead \( j+1 \). The average is calculated for all beads \( j \) adjacent to \( i \) within a designated cutoff distance, \(r_{p_2}\) (as illustrated in the supplementary material, Fig. ~S3(a). This metric helps distinguish between crystalline and non-crystalline segments in polyethylene systems \cite{10.1063/1.3608056, yi2013molecular, anwar2013crystallization, anwar2015crystallization, yamamoto2019molecular, nicholson2016analysis, hall2019divining}. These methods aim to analyze local and global ordering within the polymeric system by defining order parameters at the monomer or Kuhn segment levels, or more broadly, through a conformation tensor based on the configurations of entire chains \cite{sheng2022polymer}.

A threshold \( p_2^{\text{thr}} \) is set to determine beads within the crystalline phase. In addition, beads \( i \) and \( j \) are classified within the same crystalline domain if their interbead distance \( r_{ij} \) is below a specific threshold \( r^{\text{thr}} \). Therefore, selecting the appropriate values for \( r_{p_2} \), \( r^{\text{thr}} \), and \( p_2^{\text{thr}} \) is essential. The determination of these parameters is facilitated by analyzing the probability density functions (PDFs) of \( p_2(i) \) for various \( r_{p_2} \) in a quenched system; a random snapshot of the system that contains both crystalline and molten phases, as shown in the supplementary material, Fig. S3(c). The \( r_{p_2} \) and \( p_2 \) that coincide with the minimum between the two PDF peaks are selected as discriminative values. In our system, the threshold values $r_{p_2} = 6.0~\AA$  and $p_2 = 0.6$ were chosen. To establish \( r^{\text{thr}} \), the radial distribution function (RDF) of the system at 300~K was used (see supplementary material, Fig. S3(b)), a cutoff distance slightly above the second distance from the unbonded neighbor in a reference crystalline structure was used; this is approximately 5.8~\AA~or $1.4 \, \sigma_{\rm{CH_2}}$.

\subsubsection{Thermodynamic-like parameters for phase transitions}\label{sec:methods_ent}

The definition of local entropy, \(\bar{S}_i\), and local enthalpy, \(\bar{H}_i\), as localized thermodynamic-like quantities reflected in the configurational and energetic states surrounding a specific atom $i$, has proven effective in differentiating between liquid and various crystal symmetries \cite{nettleton1958expression, piaggi2017enhancing, piaggi2017entropy}. Local enthalpy is straightforward to compute by decomposing the interatomic potential into individual atomic energies as
\begin{equation}
    H_i = U_i + \frac{pV}{N} ~,
    \label{eq:atomicenthalpy}
\end{equation}
where \(p\) and \(V\) are the pressure and volume of the system, respectively, and \(N\) is the total number of atoms, with \(U_i\) representing the potential energy of the atom \(i\). The average local enthalpy for the atom $i$ in the defined neighbor list of the total number of atoms $N_{\text{neigh}}$ can be calculated as
\begin{equation}
    \bar{H}_i = \frac{\sum_j H_j + H_i}{N_{\text{neigh}} + 1} ~.
    \label{eq:averageatomicenthalpy}
\end{equation}

On the other hand, the local entropy calculation is more complicated and involves configurational probability distributions. For crystallization processes, an approximation using the expansion of configurational entropy, primarily the two-body excess entropy derived from Kirkwood's multiparticle correlation expansion: $S = S_{ideal\ gas} + S_2 + S_3 +\ \ldots$, is used. This approximation, which predominantly uses the pair correlation function, accounts for about $90\%$ of the configurational entropy \cite{baranyai1989direct, laird1992calculation, piaggi2017enhancing, tanaka2019revealing} and has been valuable in studies of disorder-disorder phenomena, dating back to the foundational work of Kikuchi \cite{kikuchi1951theory}. This term is expressed as
\begin{equation}
S_2 = - \, 2\pi\rho k_B \int_0^\infty [g(r) \ln g(r) - g(r) + 1] r^2 \, dr ~,
    \label{eq:atomicentropy}
\end{equation}
where \(\rho\) is the density. The local configurational entropy, \(S_i\), of the atom \(i\) is calculated using a smoothed radial distribution function,
\begin{equation}
    S_i = -2 \pi \rho_N k_B \int_0^{r_m} [ g^i(r) \ln g^i(r) - g^i(r) + 1] r^2 \, dr ~,
    \label{eq:localentropy}
\end{equation}
where \(\rho_N\) is the density of the total number of particles, \(r\) is the spatial coordinate, and \(g^i(r)\) is the modified radial distribution function for the \(i\)-th particle, with a cutoff distance, \(r_m\), given by
\begin{equation}
g^i_m(r) = \frac{1}{4\pi\rho_Nr^2} \sum_j \frac{1}{\sqrt{2\pi\sigma^2}} e^{-(r-r_{ij}^2)/(2\sigma^2)} ~.
    \label{eq:cutoof}
\end{equation}
Here, \(r_{ij}\) represents the distance between the atoms \(i\) and \(j\) that do not merge within the cutoff distance, and \(\sigma\) is a broadening parameter. This method provides an instantaneous entropic snapshot of the system at each position within the simulation cell, which can be tracked over time. To further refine the distinction between phases, the average local entropy \(\bar{S}_i\) is defined as
\begin{equation} \label{eq:Sbar}
\bar{S}_i = \frac{\sum_j S_j f(r_{ij}) + S_i}{\sum_j f(r_{ij}) + 1} ~,
\end{equation}
where \( f(r_{ij}) \) is a switching function that smoothly decays from 1 to 0 as \( r_{ij} \) approaches \( r_c \), shifting from 1 for \( r_{ij} \ll r_c \) to 0 for \( r_{ij} \gg r_c \). Piaggi et al. \cite{piaggi2017entropy} employed the following function:
\begin{equation}
f(r_{ij}) = \frac{1 - (r_{ij}/r_c)^N}{1 - (r_{ij}/r_c)^M},
\end{equation}
with \( N=6 \) and \( M=12 \). Alternatively, Nafar et al. \cite{nafar2020thermodynamically} simplified this function to:
\begin{equation}
f(r_{ij}) = 
\left\{
\begin{array}{ll}
1 & \text{if } r_{ij} \leq r_c, \\
0 & \text{if } r_{ij} > r_c.
\end{array}
\right.
\end{equation}

Previous research \cite{nafar2020thermodynamically, nafar2021communication, edwards2022nonequilibrium, edwards2023a, edwards2024a, Baig2010a, Baig2010b} has demonstrated that thermodynamic-like variables can effectively capture the complex spatio-temporal dynamics involved in nucleation and growth processes in polymeric systems. Furthermore, tuning parameters such as sigma, the cut-off distance for the system, and the discussion of local density versus global density, as discussed by \citet{nafar2020thermodynamically}, warrant careful consideration. Building on this foundation, the current study employs a thermodynamically similar parameter to determine the entropy threshold for crystallization in the simulation design. To identify an optimal entropy threshold, the system was analyzed in various states: a fully melted system, a partially crystallized system, and a fully crystallized solid system, as illustrated in the supplementary material, Fig. S5(a). The plots of the probability distribution function indicate that the optimal threshold for distinguishing between the solid and liquid states in the system is $\bar{S}_i = - \, 5.8$ (dimensionless LJ units) for the chosen model parameters, the same value used by Nafar Sefiddashti et al. \cite{nafar2020thermodynamically}.

To enhance the resolution of local entropy values and capture finer configurational details, we also briefly considered entropy values computed within discrete radial shells (\(S_{b1}, S_{b2}, \ldots, S_{b6}\)), termed \emph{entropy bands}, which represent averaged entropy within concentric neighborhoods around each atom. These were used only in the initial dimensionality reduction step (UMAP) to increase representational richness. However, to avoid biasing subsequent supervised analyses and the crystallinity index ($C$-index) development towards entropy-based descriptors, these entropy bands were excluded from all classification tasks and the final $C$-index definition. A detailed introduction and comprehensive analysis of entropy bands, alongside comparisons to spectrum-based structural descriptors such as Smooth overlap of atomic positions (SOAP) \cite{bartok2013representing, de2016comparing}, are presented separately in our forthcoming paper.

\subsubsection{Bond-orientational order (BOO) parameters}

The local symmetries of the atomic environment can be quantified using Steinhardt parameters of the bond orientation order $q_{lm}(i)$ \cite{steinhardt1983bond}. 
These parameters quantify the angular orientation between vectors (conceptually represented as bonds) joining a central particle to its neighbors. In a spherical coordinate system, each particle $i$ is considered by angles $\theta$ and $\phi$ with respect to a neighboring particle $j$ within a specified cut-off distance. The vector $\vec{R}_{ij}$ spans particles $i$ and $j$. Each of these directions between particles $i$ and $j$ can be used to associate each vector with spherical harmonics $Y_{lm}(\theta_{ij}, \phi_{ij})$. The desired symmetry in the system can dictate the choice of $l$. The values of $Y_{lm}$ are typically averaged over the first nearest neighbors of the particle $i$, or with additional information from the second shell \cite{lechner2008accurate} to compute the vector $q_{lm}(i)$, as
\begin{equation}
q_{lm}(i) = \frac{1}{N_b(i)} \sum_{j=1}^{N_b(i)} Y_{lm}(\theta_{ij}, \phi_{ij})
~.
\label{eq:qdef}
\end{equation}
In this expression, $N_b(i)$ is the number of nearest neighbors of the particle $i$. Upon averaging $q_{lm}(i)$ with neighbors:
\begin{equation} \label{eq:avg_over_neighbor}
	\bar{q}_{lm}(i) = \frac{1}{\tilde{N}_b(i)} \sum_{k=0}^{\tilde{N}_b(i)} q_{lm}(k) ~,
\end{equation}
where \( \tilde{N}_b(i) \) reflecting the number of neighbors that include the particle \( i \), the locally averaged bond order parameter \( \bar{q}_l(i) \) is calculated by summing over all harmonics,
\begin{equation} \label{eq:sum_over_l}
	\bar{q}_l(i) = \sqrt{\frac{4\pi}{2l+1} \sum_{m=-l}^{l} |\bar{q}_{lm}(i)|^2} ~.
\end{equation}

These BOO parameters are rotationally invariant and hence suitable for identifying the directional ordering in crystalline regions; yet their effectiveness is influenced by how neighbor lists are defined. 
The descriptor employed here incorporates data from the second shell of neighbors, termed averaged local bond order parameters (ALBO) \cite{lechner2008accurate} and implemented in Python Structural Environment Calculator (PYSCAL3) \cite{menon2019pyscal}. 
The Voronoi-weighted Steinhardt parameters, as proposed by Mickel et al. \cite{mickel2013shortcomings}, were also calculated but were found to be less effective in more complex polymer crystallization processes. 
This limitation arises because the number of faces in Voronoi tessellation is not particularly informative for polymer crystals in contrast to that of small-molecule crystals. However, future research could explore alternative advanced neighbor selection heuristics, such as those based on the Self-Adjusting Nearest Neighbor (SANN) algorithm \cite{van2012parameter}.

\subsubsection{Feature construction and preprocessing}
\label{sec:feature_preprocessing}
The local environment of each particle is characterized by a comprehensive feature vector constructed from multiple descriptors. Formally, the characteristic vector for a particle is defined as $\vec{X} = (\vec{X}_{OP1}, \vec{X}_{OP2}, \ldots)$, where each component $\vec{X}_{OPi}$ represents a specific order parameter (e.g., a Steinhardt value $Q_l$ or a thermodynamics-based metric). The overall dimension of $\vec{X}$ thus depends on how many descriptors are concatenated. Consequently, for $N$ particles at a given time step, the full feature space can be expressed as $N \times dim(\vec X)$. The exact list of the descriptors and their dimensions is provided in the relevant sections of \S~\ref{sec:RandD}; when not specified, the full set of order parameters is used.

After constructing the feature vectors, certain preprocessing steps were performed to prepare the data set for subsequent analyses. Continuous characteristics were standardized \textbf{} to have zero mean and unit variance, to balance their contributions during dimensional reduction and grouping steps. There were few \textbf{missing data} in chain ends for $p_2$. We imputed $p_2$ to the lower quartile as chain ends are unlikely to participate in crystal formation. We did not use missing indicator columns. Another preprocessing step was done after clustering, where the output was discretized into two primary classes, crystalline--like and amorphous--like, allowing for straightforward supervised classification and evaluation.

\subsection{Overview of machine learning workflow}

The analysis involved a structured sequential ML workflow consisting of four main stages. Initially, comprehensive descriptors that capture geometric, thermodynamic-like, and symmetry-based characteristics were calculated and pre-processed. Next, dimensionality reduction methods, namely, Principal Component Analysis (PCA)~\cite{hotelling1933analysis},  Uniform Manifold Approximation and Projection (UMAP)~\cite{mcinnes2018umap}, and variational autoencoders (VAE)~\cite{kingma2013auto} were used to compress high-dimensional feature spaces into informative low-dimensional embeddings. After evaluations, the UMAP embedding and the three-layer Variational Auto Encorder (VAE) could be used to achieve high-quality embeddings.
The results were clustered using HDBSCAN~\cite{mcinnes2017hdbscan}, Gaussian Mixture Models (GMM)\cite{day1969estimating}, and K-Means\cite{macqueen1967some} to identify different structural states.
After selecting the best combination and the clustering results, binary high-fidelity label data is generated, with known error rate, and supervised classifiers (logistic regression, random forest, gradient boosting) are used as shortcuts for model explanation, and feature importance analyses guided the construction of a robust and parsimonious crystallinity index ($C$-index).

\subsection{Low-dimensional embedding}\label{sec:umap_methods}

High-dimensional descriptor spaces impede clustering due to both computational cost and interpretative complexity. Three-dimensionality reduction techniques, PCA, UMAP, and VAE, were applied to the high-dimensional per-particle feature data to reveal latent structures associated with crystallization. After evaluations, the UMAP embedding and the three-layer VAE achieved high-quality embeddings.
After manual evaluations in 3D spatial space, clarity of separations in lower-dimensional embedding space, and considering computational cost, UMAP was adopted for most of the subsequent analyses. 
The hyper parameters of UMAP were tuned by grid search: the number of neighbors was varied across a practical range to balance local detail against noise smoothing, while the minimum distance parameter was kept near zero to sharpen the separation between clusters, as detailed in \S~\ref{sec:clust_methods}.

\subsection{Unsupervised clustering of UMAP-reduced feature space} \label{sec:clust_methods}

Clustering was performed on low-dimensional embeddings obtained from UMAP. Specifically, embeddings reduced to two dimensions were analyzed using HDBSCAN, GMM, and K-Means algorithms to identify distinct structural states. 

HDBSCAN was selected over DBSCAN for its ability to detect clusters of varying densities without requiring predefined cluster counts or global distance parameters.
Its hyperparameters, \texttt{min\_cluster\_size} and \texttt{min\_samples}, were systematically optimized at multiple simulation time steps; tuning details are provided below. Due to changes in structural organization over time, these hyperparameters required re-tuning across different temporal regimes to maintain clustering accuracy. The slight variation observed in optimal values between time steps highlights this necessity, consistent with previous findings~\cite{adorf2019analysis}.

GMM provided complementary cluster characterization by modeling the data as a mixture of Gaussian distributions optimized through expectation-maximization, producing smooth probabilistic boundaries. K-Means served as a simpler baseline for cluster assignment comparisons.

\subsection{Supervised model validation and feature selection} \label{sec:classification_methods}

Although clustering provides valuable structural identification, relating discovered clusters back to high-dimensional descriptors directly is computationally expensive and challenging to interpret. To address this, supervised classifiers—Logistic Regression, Random Forest and Gradient Boosting (implemented through \texttt{ scikit-learn}~\cite{pedregosa2011scikit}) were trained on the original descriptors using cluster labels from the unsupervised stage as targets. Data were split into training and test sets using an 80/20 ratio, unless otherwise stated. The performance of these models was evaluated using precision, recall, and the area under the ROC curve (AUC).

Feature importance analysis was used to interpret the results.
For logistic regression, standardized coefficients were used as proxy and for the tree-based model, permutation importance was used. We also employed SHAP (SHapley Additive exPlanations)~\cite{NIPS2017_7062, lundberg2018explainable}, which attributes feature contributions to individual predictions, enabling localized interpretation of the ML model.
This allowed us to investigate how each of the features contribute to the classification of atoms, with a focus on ambiguous atoms close to cluster boundaries.
Lastly, conditional independence via Fisher's Z test guided the selection of optimal low-dimensional subset of features and the construction of the $C$-index as a versatile OP for polymer crystallization.

\section{Results and Discussion}
\label{sec:RandD}

\subsection{Molecular dynamics simulation and initial observations}
\label{sec:31_density_vs_t}

\begin{figure}[h!]
\centering
\includegraphics[width=\textwidth]{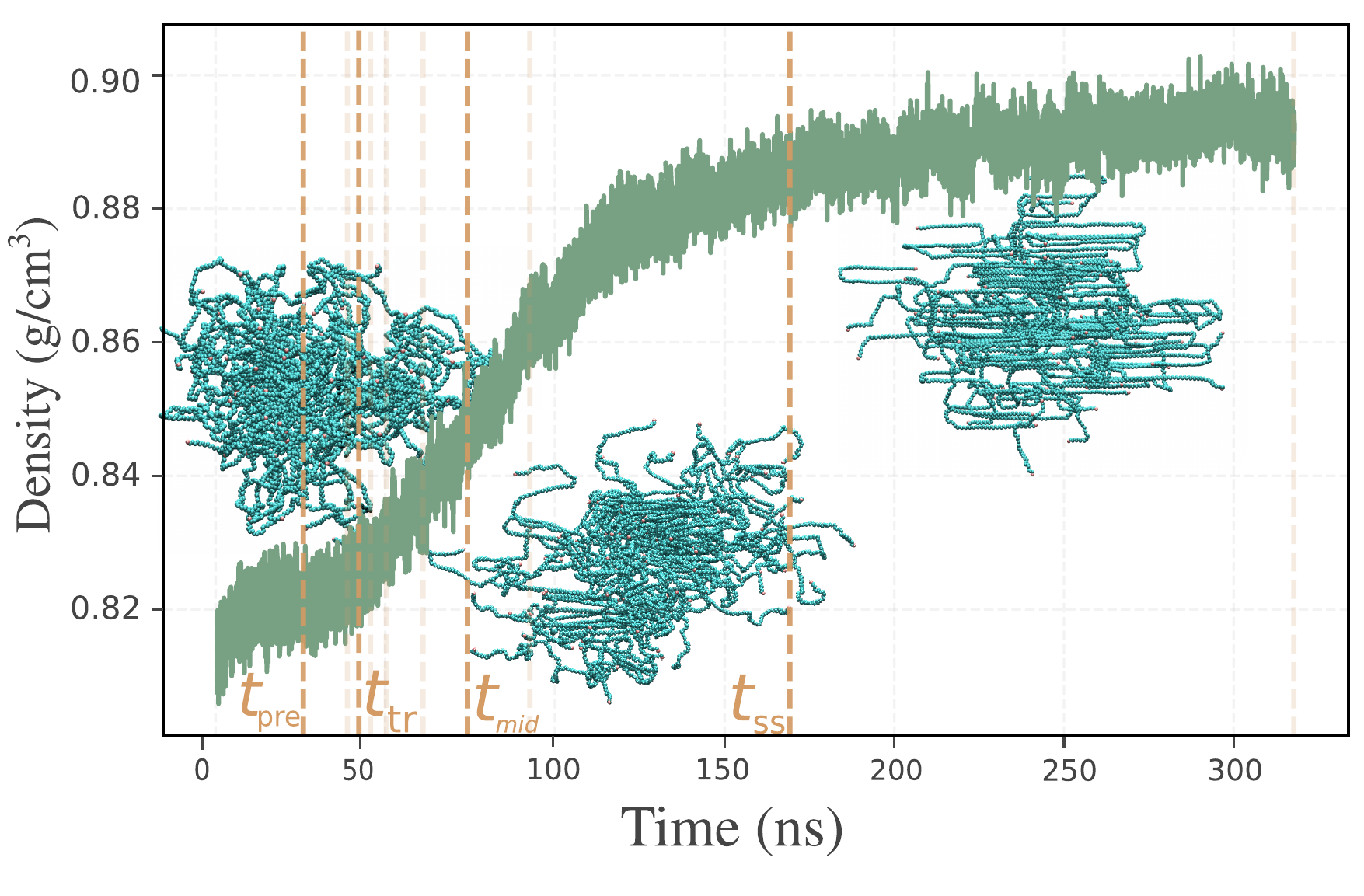}
\caption{
Density evolution during rapid quenching to 300~K shows the transition from isotropic melt to semicrystalline state. Vertical orange dashed lines mark four key snapshots ($t_{pre}$, $t_{tr}$, $t_{mid}$, $t_{ss}$) used for analysis; lighter dashed lines indicate additional points for ML training. VMD snapshots illustrate structural changes during nucleation and growth.
}
\label{fig:timeDensity}
\end{figure}

The self-assembly of the model polymer chains was investigated as described in \S\ref{sec:methods_md}. We analyze ten snapshots sampled along the quenching trajectory, shown by the vertical dashed lines in Figure~\ref{fig:timeDensity}.
The system was rapidly cooled until its density stabilized and no further
significant changes in the cluster structure were observed. The timing of these snapshots is shown in Figure~\ref{fig:timeDensity}, where each bold orange dashed line mark a key point in the crystallization process
used for analyses in the following sections:
{\textbf{$t_{pre}$}}, an early stage where small nuclei sporadically form and dissolve;
{\textbf{$t_{tr}$}}, a transition point where stable nuclei begin to grow;
{\textbf{$t_{mid}$}}, the midpoint of the crystallization process; and
{\textbf{$t_{ss}$}}, a steady-state regime where crystalline domains coexist with amorphous regions.
This selection ensures consistency with previous nucleation studies
\cite{bhardwaj2024nucleation} and provides a structured basis for exploratory and quantitative analysis.

The insets of the VMD~\cite{humphrey1996vmd} snapshots in Figure~\ref{fig:timeDensity} illustrate the critical stages during polymer crystallization in unwrapped coordinates.
In the initial snapshot \(t_{pre}\), the polymer chains exhibit an isotropic and highly disordered configuration typical of the molten state prior to nucleation.
By the midpoint snapshot \(t_{mid}\), a distinct nucleation cluster has emerged, characterized by local chain alignment and partial ordering.
The transition snapshot at $t_{tr}$ is not shown, as its visual structure closely resembles that of $t_{pre}$ at this scale.
Finally, in steady state (\(t_{ss}\)), polymer chains form aligned crystalline domains interspersed with residual amorphous regions.

Note that in this specific simulation setup (60C150), a single stable nucleus dominates growth due to the deliberately chosen system size; however, larger systems typically show multiple independent nucleation events occurring at the same time. The presence of multiple nuclei in larger systems can result in chains aligning in multiple directions, unlike the predominantly unidirectional alignment shown here.

\subsection{Discordance Among Conventional Order Parameters}
\label{subsec:discordance}

Common-order parameters (such as local density from Voronoi analysis, \(p_2\), entropy, and \(q_l\)) each capture distinct aspects of crystallization; however, they often yield inconsistent results when identifying local crystalline environments.
For example, densely packed folded regions can display low entropy along with low \(p_2\), while surface atoms may exhibit broken symmetry in \(q_6\) despite being part of crystallized domains.

The local density based on Voronoi tessellation identifies crystalline regions in steady state (with some considerable noise) by applying a threshold derived from the reference orthorhombic structure ($\rho_{local, ortho}^{mean} - \sigma = 0.922$); see supplementary material, Sect. S1.1. However, in the early stages ($t_{pre}$), density fluctuations caused by temporary chain entanglements compromise its reliability. More dramatically, the distribution of Voronoi face counts lacks clear phase differentiation, limiting its usefulness as a crystallinity metric; see the supplementary material, Fig. S1(b).

Angle-based alignment order parameter ($p_2$) captures the emergence and growth of crystallinity effectively, identifying transient local ordering in early crystallization stages and mature crystalline domains in steady state; see the supplementary material, Fig. S4. 
Although \( p_2 \) is an intuitive and physically meaningful descriptor of local chain alignment, it has several limitations. First, it cannot be computed for chain-end atoms due to the lack of definable bond vectors. Second, \( p_2 \) values tend to fluctuate near crystal surfaces where local distortions disrupt ideal alignment, sometimes resulting in incorrect classification of surface particles. Finally, the sensitivity of \( p_2 \) to the choice of neighbor cutoff and averaging window can reduce its robustness when applied to systems with varying chain stiffness, length, or interaction potentials, and its distribution shows extensive overlap between phases.
Despite these challenges, \(p_2\) remains a useful baseline order parameter due to its low computational cost and direct physical interpretation. In later sections, it is quantitatively compared with thermodynamic-like, spherical harmonic, and machine learning-based descriptors to assess its relative performance.

\begin{figure}
    \centering
    \includegraphics[width=0.99\linewidth]{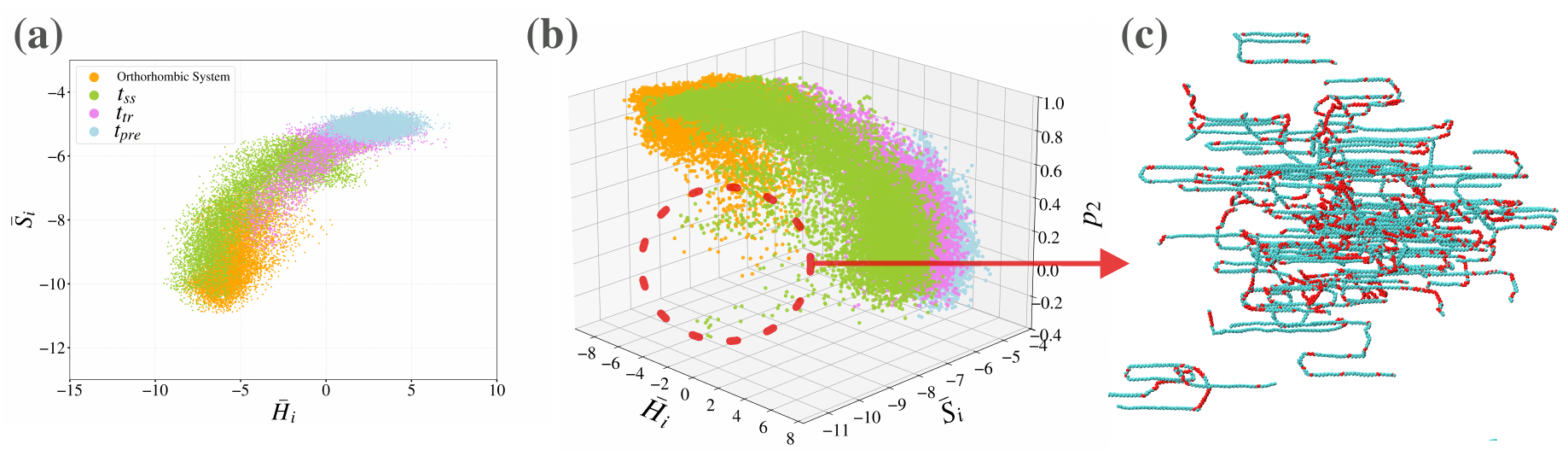}
    \caption{Joint distributions of thermodynamic-like descriptors in a quenched polymer system. (a) The 2D plot of entropy $\bar S_i$ vs. enthalpy $H_i$ shows a continuous spread without distinct phase clusters, unlike prior reports \cite{piaggi2017entropy}. (b) Adding $p_2$ in 3D space ($\bar H_i$, $\bar S_i$, $p_2$) does not improve separation; notable discrepancies appear in high-entropy, low-enthalpy regions. (c) A crystalline configuration highlights that entropy–$p_2$ disagreements occur mainly in folded and bridged segments; 
    }
    \label{fig:thermo_joint}
\end{figure}

Thermodynamic-like descriptors, such as entropy ($\bar{S}_i$) and enthalpy ($\bar{H}i$), provide an alternative means to assess structural order. Figure~\ref{fig:thermo_joint}(a) shows the joint distribution of $\bar{S}i$ and $\bar{H}i$ in key crystallization stages:$t{pre}$ (melt), $t{mid}$ (early ordering), $t{ss}$ (steady state) and the orthorhombic crystal reference. Unlike previous studies on small molecules \cite{piaggi2017enhancing}, the polymeric system shows a continuous distribution without clear cluster separation, probably due to its complex internal structures and dynamics.

Figure~\ref{fig:thermo_joint}(b) incorporates the parameter $p_2$ to form a 3D space, yet the additional dimension still fails to cleanly separate the crystalline and amorphous states. In particular, discrepancies between entropy and the values of $p_2$ persist in folded or bridged chain segments (Figure~\ref{fig:thermo_joint}(c)), where low entropy is observed alongside moderate $p_2$. This suggests that local structural complexity introduces discordance between otherwise reliable metrics.

In addition to geometric and thermodynamic-like descriptors, BOO parameters ($q_\ell$) ~\cite{steinhardt1983bond,lechner2008accurate} were also assessed. Individual scalar BOO parameters ($q_2$, $q_4$, $q_6$, $q_8$, $q_{10}$) showed some discriminative power but were limited by overlapping distributions between crystalline and amorphous phases (supplementary material, Fig. S6). Even two-dimensional projections such as the commonly used $q_4$–$q_6$ distribution ~\cite{musil2021physics, geiger2013neural, chakravarty2007lindemann} failed to produce sharply separated clusters, supporting the notion that polymer crystallization involves gradual, continuous structural transitions rather than clear phase boundaries.
This observation is consistent with the scalar entropy and $p_2$ distributions presented earlier (see the supplementary material, Fig. 5(a) and 3(c)), where broad and overlapping distributions and gradual transitions between disordered and ordered phases were observed. Together, these results underscore the absence of sharp phase boundaries in polymer crystallization and motivate the use of higher-dimensional or multi-descriptor approaches.

\begin{figure}
    \centering
    \includegraphics[width=\linewidth]{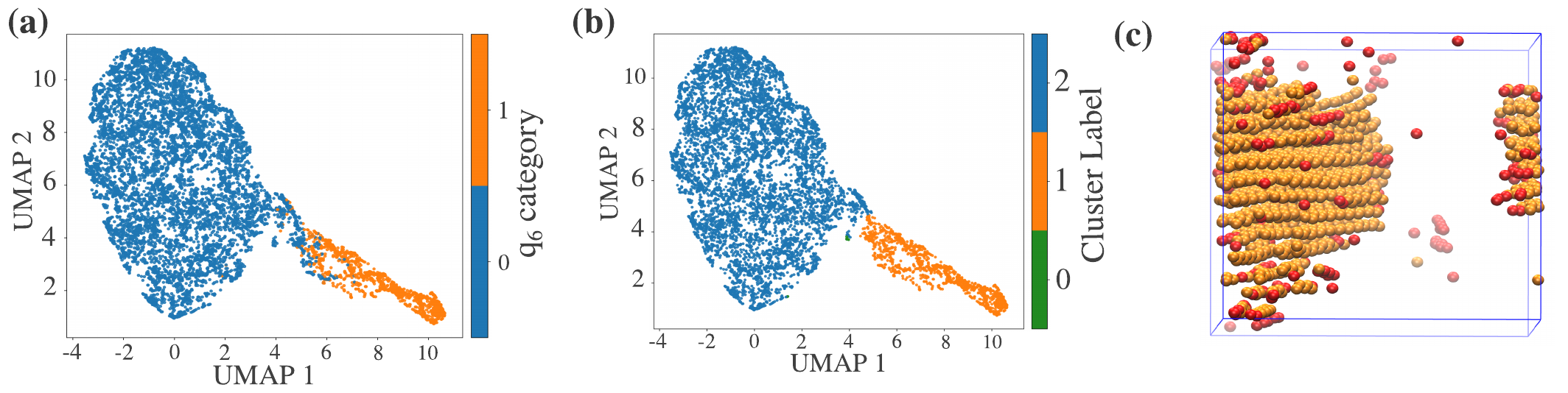}
    \caption{
    (a) 2D UMAP embedding of BOO fingerprints colored by $q_6$-based crystallinity classification. 
    (b) HDBSCAN clustering on the same embedding, showing similar phase separation but with improved boundary delineation.
    (c) 3D visualization of crystalline particles: orange particles meet the $q_6 > q_6^{thr}$ threshold, while red particles are additionally identified as crystalline by HDBSCAN but not identified by scalar $q_6$. These particles appear primarily along the periphery of the crystalline domains, suggesting potential enhanced sensitivity of the BOO fingerprint to interfacial structure.
    }
    \label{fig:boo_q6}
\end{figure}

To enhance discriminative sensitivity, BOO fingerprint vectors were constructed by concatenating higher-order spherical harmonics ($q_\ell$, with $\ell \in \{2,4,6,8,10\}$. Figure~\ref{fig:boo_q6}(a-b) illustrates dimensionality reduction via UMAP followed by clustering with HDBSCAN. Compared to scalar thresholding in panel a (we use threshold $q_6^{thr}= 0.18$, see supplementary material, Fig. S5(b)), vector-based clustering, in panel (b), exhibited improved boundary delineation, particularly near domain surfaces and structurally distorted regions (Figure~\ref{fig:boo_q6}(c)). The orange particles in panel (c) exceed the $q_6^{thr}$ threshold, while the red particles are additionally identified as crystalline by HDBSCAN but fall below the scalar threshold. These red-highlighted particles frequently appear along the radial protrusions or longitudinal edges of the crystalline domains: regions where the scalar $q_6$ yields lower values, likely reflecting local geometric distortion rather than a true absence of order. However, a clear, universally applicable boundary between phases remained elusive even with vector-based BOO descriptors.

Temporal stacking of BOO descriptors over consecutive snapshots further improved cluster separability (Supplementary Fig. S9), aligning with previous studies~\cite{adorf2019analysis}. However, while incorporating temporal stacking improves phase separability, in this study, the focus is placed on a single-timestep analysis for two main reasons. First, the aim is to identify efficient and interpretable order parameters that can be later applied to more detailed scenarios, including initial nucleation of polymers and secondary nucleation during crystal growth. These applications require high spatial and temporal resolution, sometimes down to a single snapshot. Second, given the plan to combine descriptors from geometric, symmetry-based, and thermodynamic-like categories, it is unclear whether these features change in similar timescales. So we avoid relying on temporally extended vectors for now.

Pairwise correlation analyses (detailed in the supplementary material, Figs. S6-8) confirm generally weak linear dependencies through the process time.
Although most pair of descriptors produce low coefficients of determination ($R^2$ values typically in the range of 0.1 to 0.3), certain combinations, particularly within phase-separated subsets, exhibit moderately higher correlations ($R^2 \sim 0.5-0.6$).
The inconsistencies in these descriptors prompt the use of advanced nonlinear, multivariate approaches. The testing of advanced ML techniques may reveal a low-dimensional space that better distinguishes between the amorphous and crystalline phases.

\subsection{Dimensionality reduction and clustering of combined atomic descriptors}\label{subsec:combined_dimred}

\begin{figure}[htbp]
    \centering
    \vspace{-0.6em}

    \begin{subfigure}[t]{0.3\textwidth}
        \includegraphics[width=\linewidth]{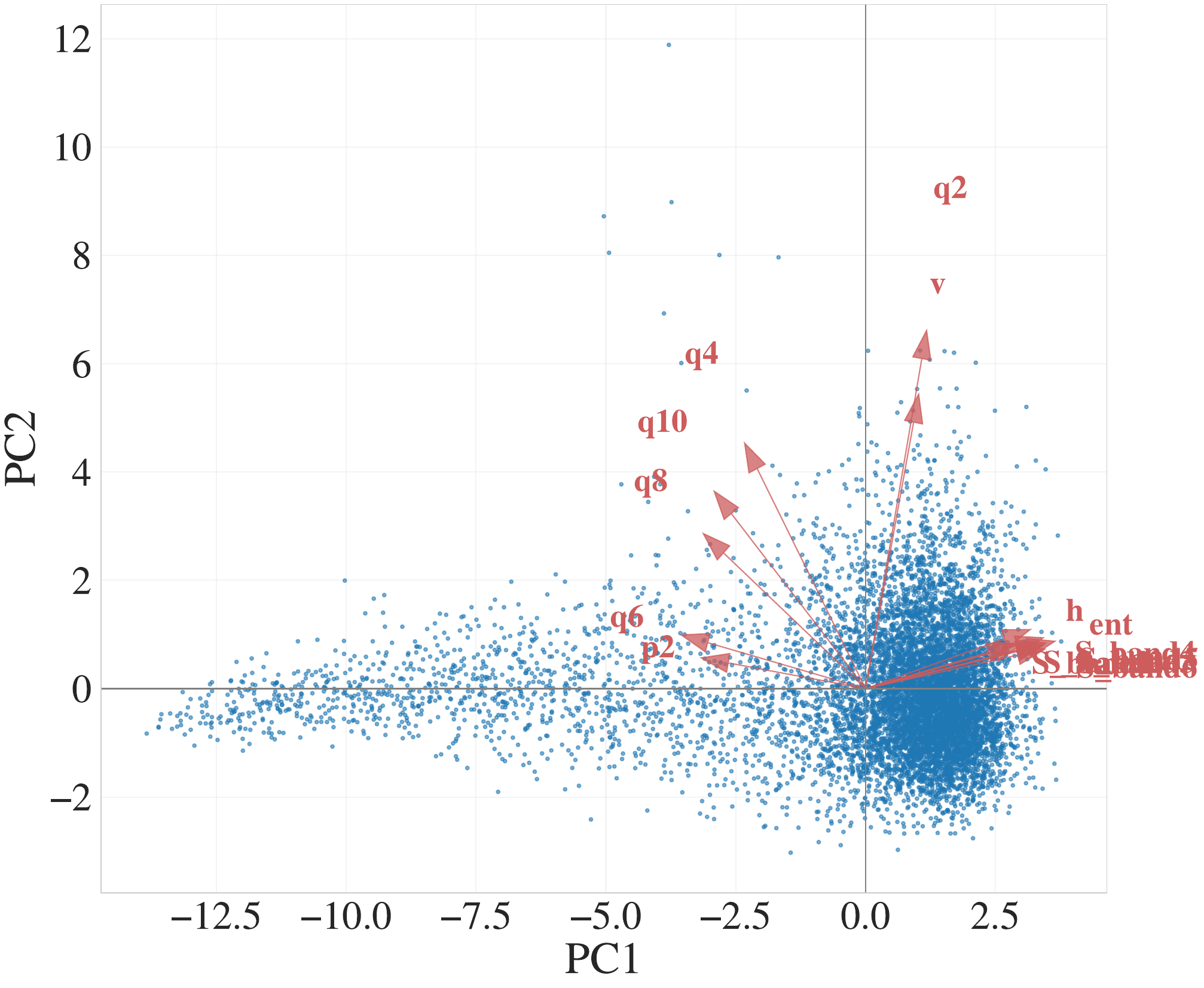}
        \caption*{\textbf{(a)}}
    \end{subfigure}
    \hfill
    \begin{subfigure}[t]{0.3\textwidth}
        \includegraphics[width=\linewidth]{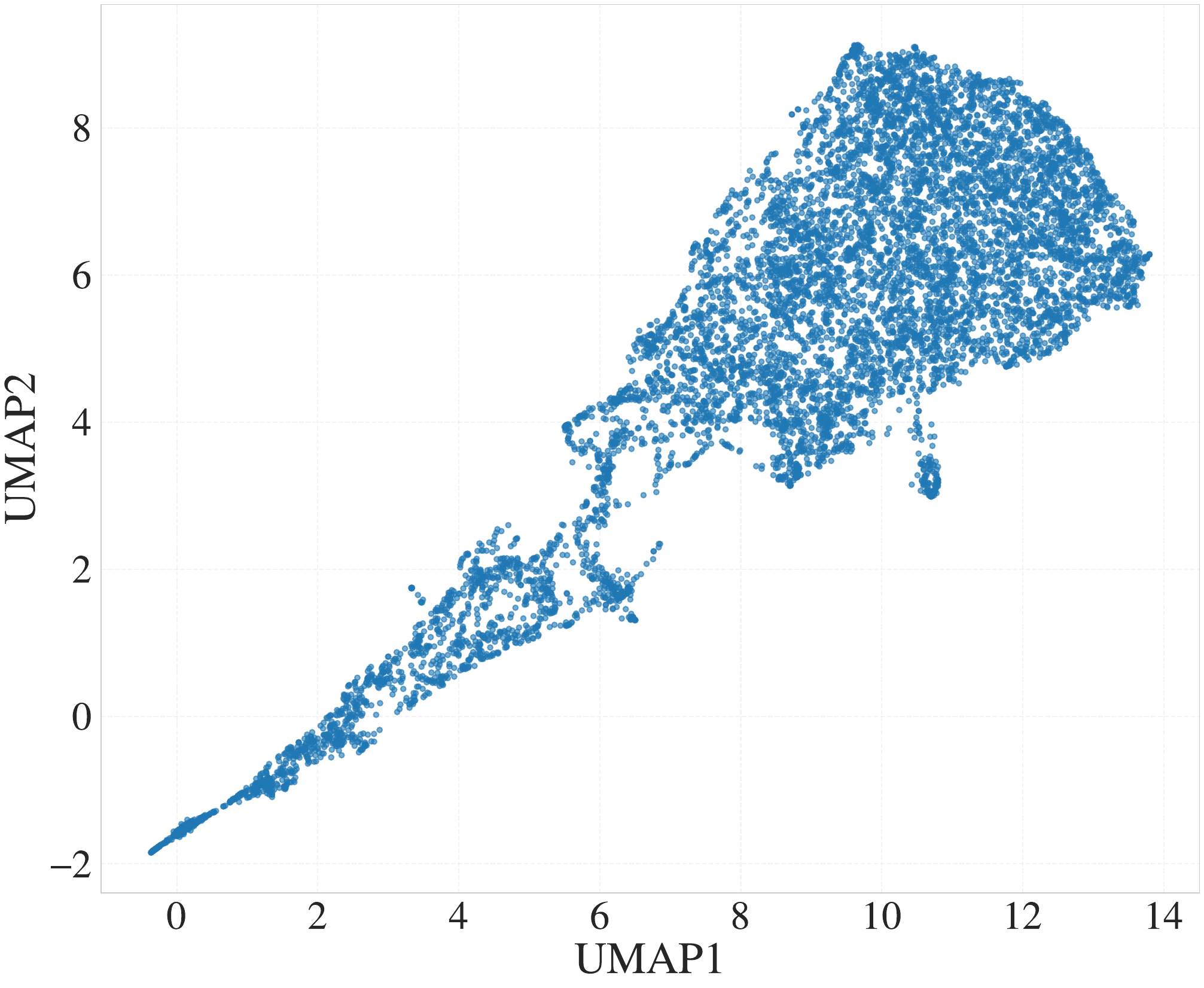}
        \caption*{\textbf{(b)}}
    \end{subfigure}
    \hfill
    \begin{subfigure}[t]{0.3\textwidth}
        \includegraphics[width=\linewidth]{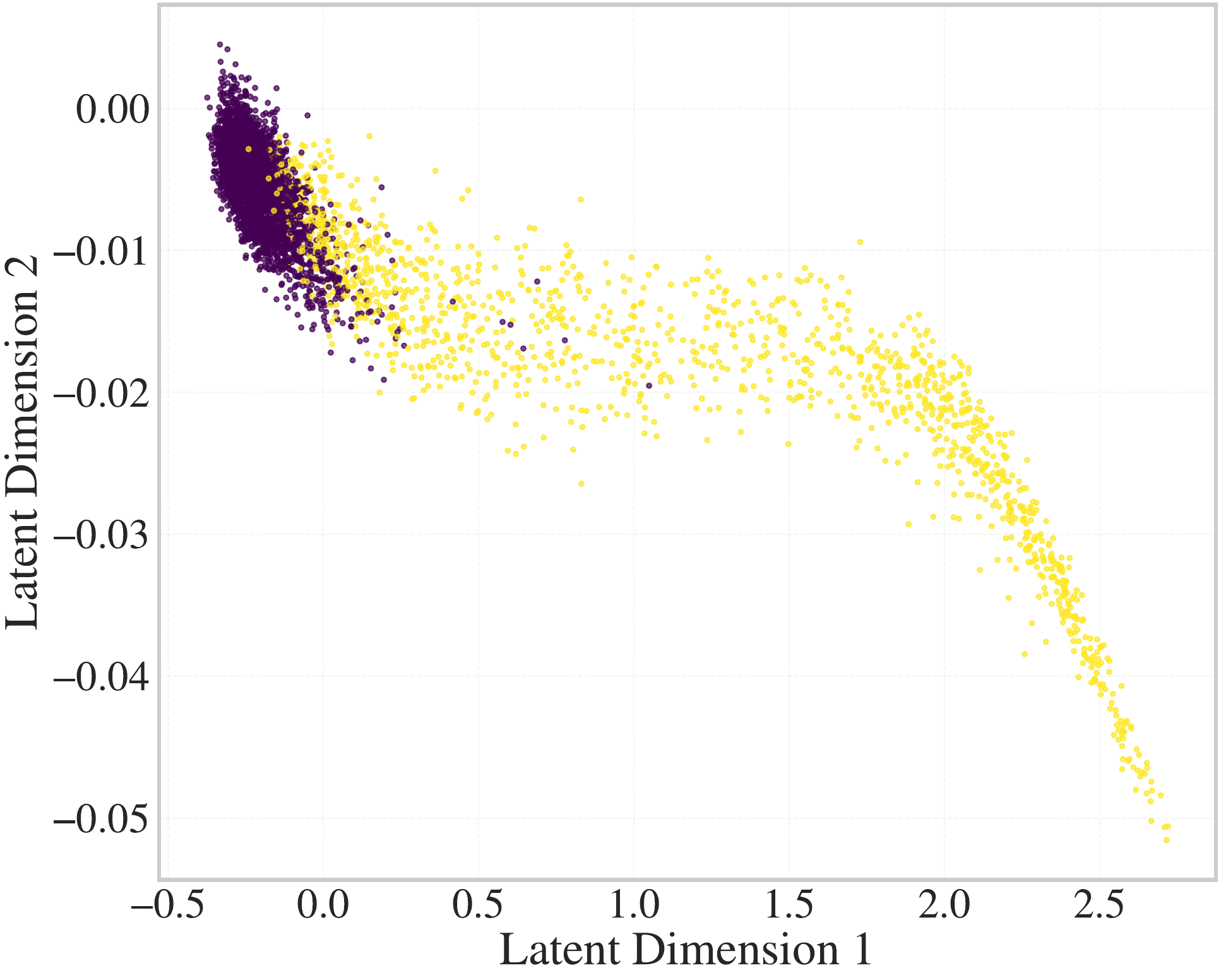}
        \caption*{\textbf{(c)}}
    \end{subfigure}

    \vspace{0.8em}

    \begin{subfigure}[t]{0.3\textwidth}
        \includegraphics[width=\linewidth]{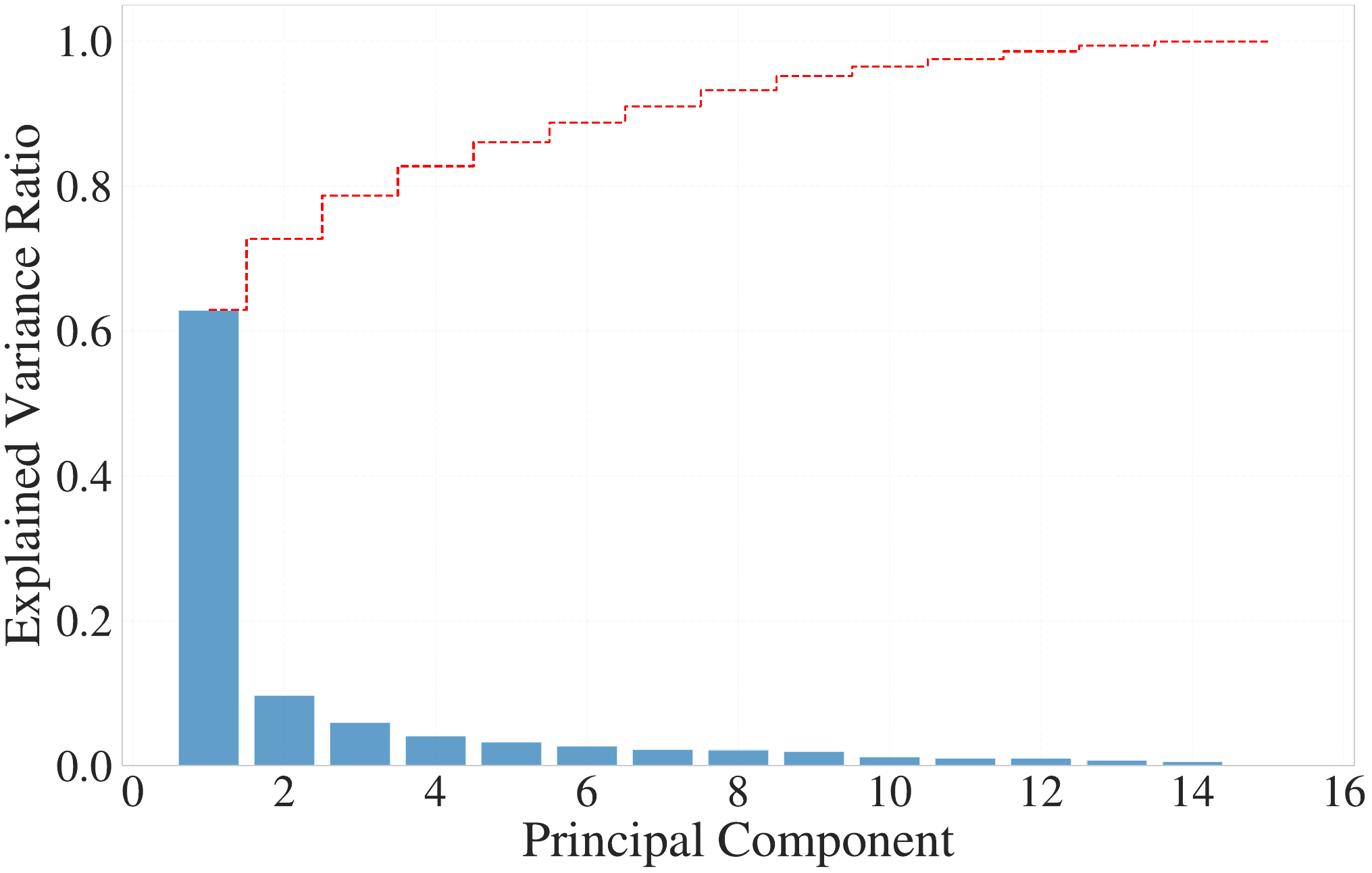}
        \caption*{\textbf{(d)}}
    \end{subfigure}
    \hfill
    \begin{subfigure}[t]{0.3\textwidth}
        \includegraphics[width=\linewidth]{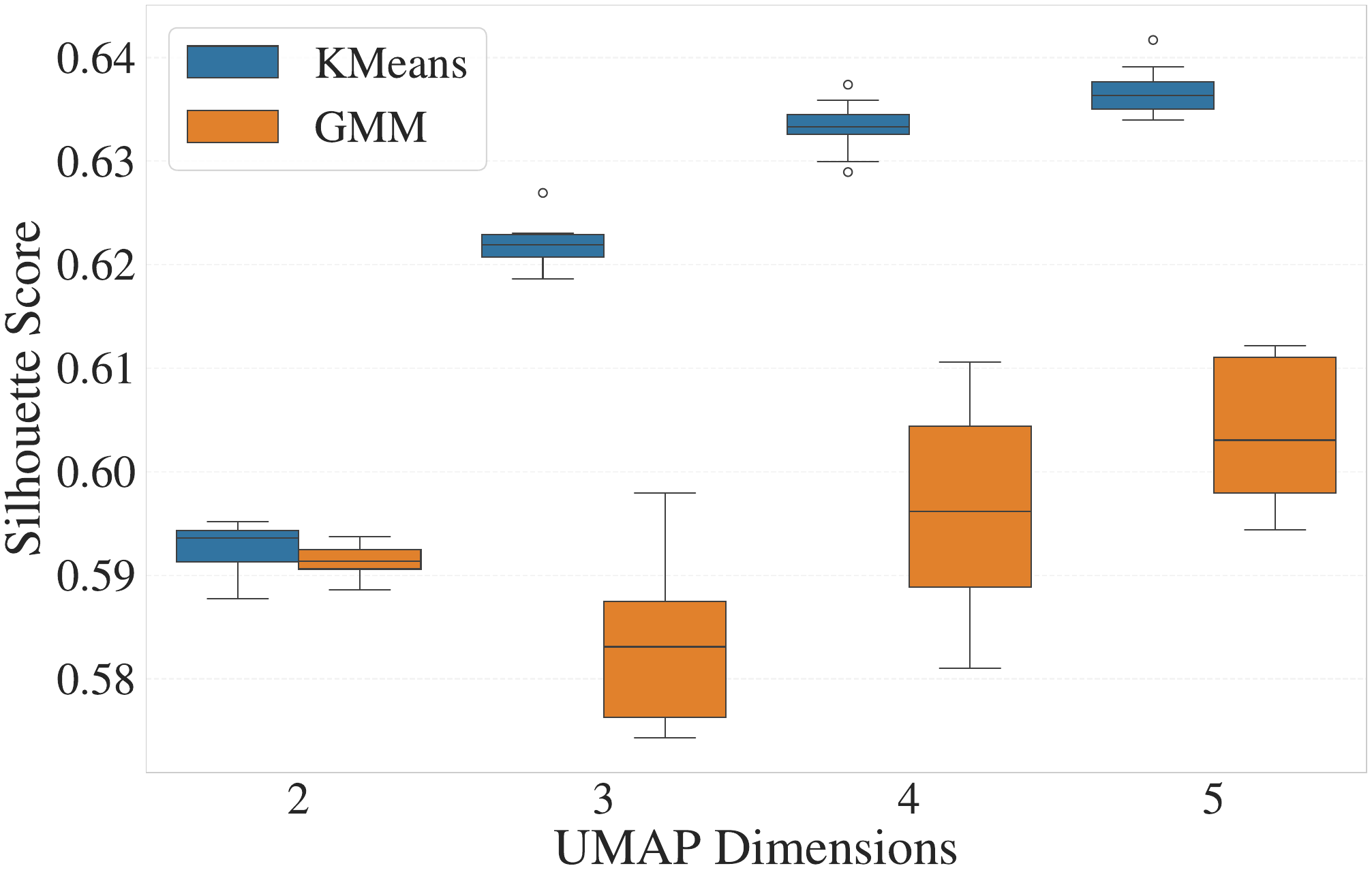}
        \caption*{\textbf{(e)}}
    \end{subfigure}
    \hfill
    \begin{subfigure}[t]{0.3\textwidth}
        \includegraphics[width=\linewidth]{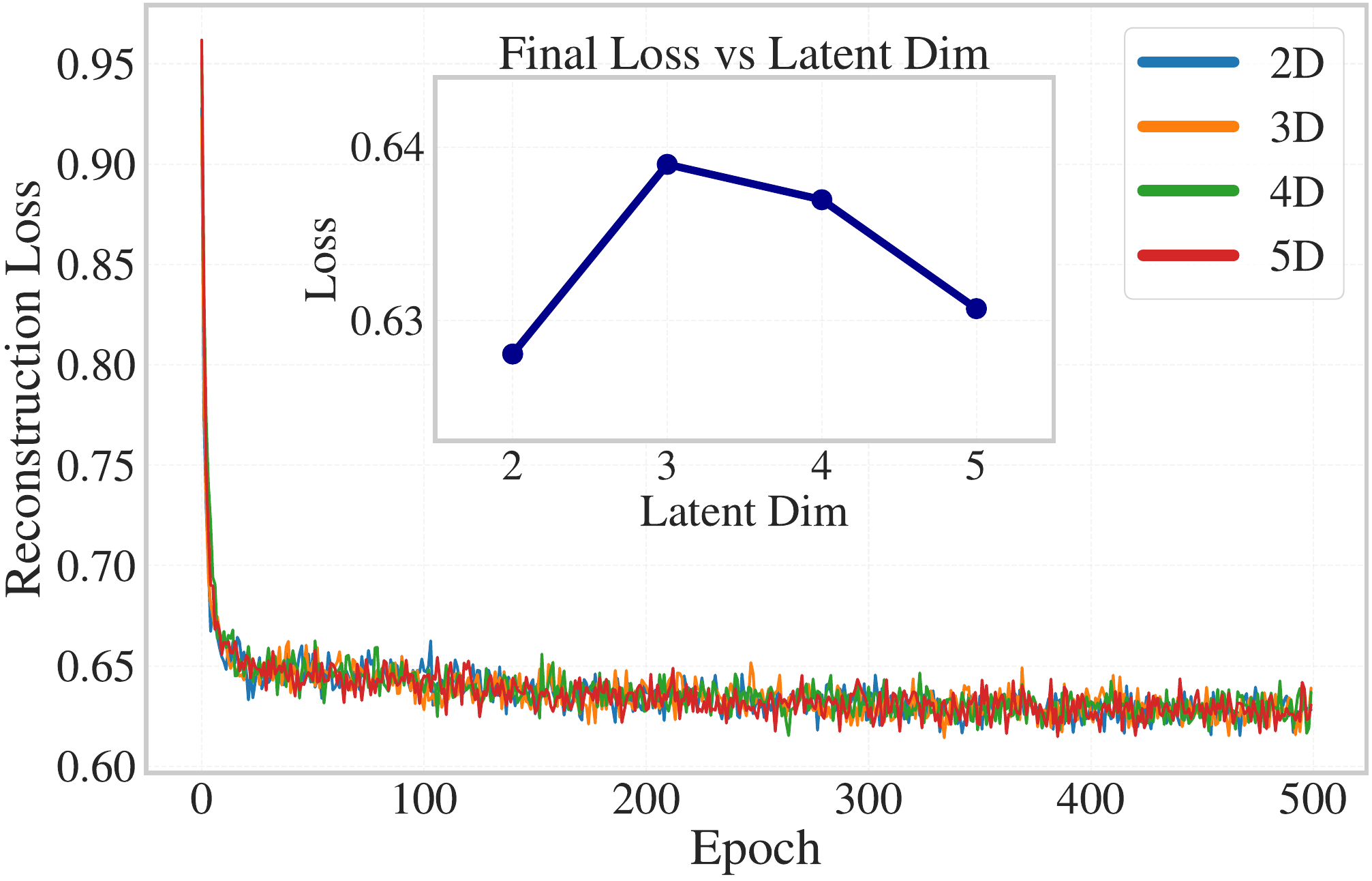}
        \caption*{\textbf{(f)}}
    \end{subfigure}

    \caption{Dimensionality reduction techniques applied to combined atomic descriptors.
    (a) PCA biplot showing projection of data onto the first two principal components with feature loadings indicated by red arrows. 
    (b) PCA variance plot showing most variance captured by first few components.  
    (c) 2D UMAP reveals nonlinear structure and separation of environments.
    (d) Silhouette scores for K-Means and GMM across UMAP dimensionalities (averaged over 10 runs). 
    (e) VAE latent space (2D) shows smooth distribution, colored by entropy-based category.
    (f) VAE training curves across dimensions; minor reconstruction differences favor 2D.}
    \label{fig:dimred}
\end{figure}

Applying PCA, the biplot in Figure~\ref{fig:dimred}(a) shows the projection of atomic environments onto the first two principal components, with red arrows indicating feature loadings. In particular, entropy, entropy bands, $p_2$, and $q_6$ emerge as the most discriminating characteristics. Note that since entropy values are negative, their corresponding arrows point in the opposite direction. The variance plot in Figure~\ref{fig:dimred}(d) confirms that only a few main components capture the majority of the variance in the data set. However, the resulting 2D projection lacks distinct separation between the structural states.
Throughout this study, PCA was used as a preprocessing step prior to UMAP only when high-dimensional Smooth Overlap of Atomic Positions (SOAP) features were combined with other descriptors. In all other cases, only UMAP was used for dimensionality reduction.

UMAP, a nonlinear manifold learning technique, produced a more stratified and informative latent space. The 2D UMAP projection (Figure~\ref{fig:dimred}(b)) reveals structural
heterogeneity, with separation between different atomic environments. This
separation was quantitatively assessed using silhouette scores, computed by
applying K-Means and GMM clustering with two clusters across increasing UMAP output dimensions.
As shown in Figure~\ref{fig:dimred}(e), silhouette scores were averaged over ten independent UMAP runs for each dimensionality setting. Although K-Means consistently achieved higher silhouette values than GMM, both methods revealed stable and interpretable trends. In particular, the best performance for K-Means was observed in 4D and 5D embeddings (up to 0.64), but the 2D embedding still maintained competitive scores around 0.6. Given its strong visual interpretation and relatively stable clustering performance, the 2D UMAP embedding was selected for subsequent clustering and visualization tasks. HDBSCAN was excluded from this comparison, as its performance was more sensitive to internal hyperparameter tuning and not directly comparable to fixed-cluster methods, such as K-Means and GMM. For all UMAP results reported in this work, the hyperparameters \texttt{n\_neighbors} = 10 and \texttt{min\_dist} = 0.0 with the \texttt{Manhattan} metric were used, as they yielded the highest silhouette scores in both K-Means and GMM clustering methods.
For more details on UMAP parameter selection, see the supplementary material, Fig. S10.

We also use VAE to learn latent representations optimized for accurate data
reconstruction.
The resulting 2D latent space, shown in Figure~\ref{fig:dimred}(c), reveals a smooth and continuous distribution of atomic environments, without sharp
separation between structural phases. Performance evaluation was carried out
using standard VAE losses.
As illustrated in Figure~\ref{fig:dimred}(f), the reconstruction loss remains relatively consistent across different latent dimensionalities. The inset highlights minor variations in loss with increasing latent dimension,
indicating stabilization beginning at 2D and persisting through higher
dimensions within the same range. This plot corresponds to a shallow encoder
architecture with a single hidden layer of 64 neurons.
To aid in interpretation, we colored the VAE projection using entropy-based
crystallinity categories because phase distinctions were not visually apparent with uniform coloring.
The capacity of VAE is further assessed by deeper architectures (see the supplementary material, Fig. S11); although there were improvements for three-layer VAE, the latent space still exhibited a weaker phase separation compared to UMAP.
(cf. Figures~\ref{fig:dimred}(b) and~(c) or supplementary material, Fig. S11(c)).
The combination of clear visual separation and consistently high silhouette
scores for UMAP, and the higher computational cost of VAE with three layers
led us to select UMAP as the primary dimensionality reduction
method for subsequent cluster analysis.

We also tested different clustering algorithms, K-Means, GMM, and HDBSCAN, on UMAP output data.
The details of performance comparison are included in the supplementary material, Sect. S3.1.
In summary, studying the boundary atoms and assessing the phase separation quality, we concluded that HDBSCAN provided the most faithful reconstruction (supplementary material, Fig. S13).
Therefore, to classify local environments as crystalline or amorphous, the primary clustering analysis was performed using HDBSCAN, applied to the UMAP-reduced feature space.

\paragraph{Label Integration and Downstream Use.}

HDBSCAN assignments were adopted as the canonical \texttt{Cluster\_Label} and appended to the atomic feature table. The identified boundaries define an uncertainty band that is used in the analyzes in Section \ref{sec:uncertainty}. These labels, together with the entire feature set, underpin supervised modeling and SHAP analysis in Section \ref{sec:shap}, where we quantify how
different order parameters modulate crystallinity in precisely the identified interfacial regions.

\subsection{Model explanation via supervised classification}\label{subsec:supervised}

\paragraph{Motivation and strategy.}
To learn how the UMAP space plus the selected clustering relates back to the
high-dimensional initial space of the OPs,
we adopt a shortcut strategy.
The generated labels data are used to relearn an easier and possibly more
generalizable version of this relationship.
And after studying its general applicability in time,
we conclude that the relearned model is a decent approximation to our cluster labels, thus creating a new crystallinity index $C$-index that essentially considers all these dimensions.

Note that using a tree-based \textit{universal approximator}, such as the random
forest model, a priori guarantees that we get a reasonable fit that matches the results
of UMAP plus clustering.
The goals here are
(i) to fit and use these models for easier model interpretation,
(ii) to learn whether a linear model can achieve acceptable performance,
(iii) to learn whether this relearned model is generalizable in time, and
(iv) if possible to simplify this relearned model.

The workflow we introduce in this paper, namely (i) using dimensional
reduction, (ii) using clustering and studying the errors, (iii) creating the
most reasonable labels data in the smaller space, and (iv) relearning the
function as an index, i.e., a much more accurate OP for the specific problem at hand, can be generalized not only to nucleation or crystallization but also in other similar systems. However, one has to deal with many OPs that represent a certain phenomenon but each have their own systematic biases.  We argue that applying this workflow only a few times along the system trajectory is sufficient to obtain a meaningful index parameter, keeping the overall computational cost low.

\paragraph{Data Preparation and Training Protocol.}
The characteristic matrix included geometrical descriptors based on Voronoi (volume, number of faces), local thermodynamic metrics (entropy, enthalpy), and bond orientation order parameters ($p_2$, $q_\ell$). All continuous features were standardized to zero mean and unit variance. The data set was divided into training subsets (80\%) and test subsets (20\%) using stratified sampling to preserve class proportions. The hyperparameters were optimized by cross-validation of the training data, maximizing the mean \textit{Area Under the Receiver Operating Characteristic Curve} (AUC) using grid search.

\paragraph{Hyperparameter Optimization.}
\textbf{Logistic Regression:} Regularized via both $L_1$ and $L_2$ penalties, looking for inverse regularization strengths $C \in [{0.01, 0.1, 1, 10}]$ and penalty types ${\ell_1, \ell_2}$.
\textbf{Random Forest:} Varied the number of trees $n_{\text{estimators}} \in [{100, 200, 500}]$ and the maximum depth of the trees $\text{max\_depth} \in [{5, 10, 20}]$, with minimum samples per leaf fixed at 1 to capture fine-grained distinctions.
\textbf{Gradient Boosting:} Searched over $n_{\text{estimators}} \in [{100, 200, 500}]$, $\text{max\_depth} \in [{3, 5, 7}]$, and learning rates $\in [{0.01, 0.1, 0.2}]$, balancing the complexity of the model against the risk of overfitting.

All grids were evaluated using 5-fold stratified cross-validation, with the mean AUC as the scoring metric. In the supplementary material, Fig. S14(a) and (b) display the AUC heat maps resulting from these grid searches for random forest and gradient boost, respectively. The relatively smooth and small variation of AUC in hyperparameter space confirms the robustness of the model, as the performance remains high (\mbox{AUC}~$\gtrsim 0.96$) across a wide range of depths and tree counts.

\paragraph{Evaluation Metrics and Test Set Performance.}
After hyperparameter tuning, each model was retrained on the entire training
set using the best-found parameters and evaluated on the excluded test set.
Table~\ref{tab:supervised_results} summarizes the final performance of the test in terms of precision, recall, and AUC. \emph{Gradient Boosting} achieved the highest overall performance with accuracy 0.9927~$\pm$0.0012, precision 0.9776$\pm$0.0078, recall 0.9731$\pm$0.0068, and AUC
0.9846$\pm$0.0032.
\emph{Random Forest} also performed well (AUC$=0.9791$), while
\emph{Logistic Regression} remained competitive (AUC~$\approx0.97$).
\begin{table}
    \centering
    \begin{tabular}{lc|c|c|c}
        \toprule
        Model & Accuracy & Precision & Recall & AUC \\
        \midrule
        Logistic Regression & 0.9871 $\pm$ 0.0014 & 0.9629 $\pm$ 0.0033 & 0.9500 $\pm$ 0.0112 & 0.9718 $\pm$ 0.0054 \\
        Random Forest & 0.9912 $\pm$ 0.0018 & {0.9788} $\pm$ 0.0064 & 0.9619 $\pm$ 0.0119 & 0.9791 $\pm$ 0.0059 \\
        Gradient Boosting & {0.9927} $\pm$ 0.0012 & 0.9776 $\pm$ 0.0078 & {0.9731} $\pm$ 0.0068 & \textbf{0.9846} $\pm$ 0.0032 \\
        \bottomrule
    \end{tabular}
    \caption{Classification performance metrics for supervised models evaluated via five-fold cross-validation.}
    \label{tab:supervised_results}
\end{table}

\paragraph{Uncertainty of \texttt{Cluster\_Label}.}\label{sec:uncertainty}
Manual inspection of 400 atoms selected near the UMAP cluster boundary at two time steps (200 boundary atoms for each timestep) revealed that 6 atoms (1. 5\%) were clearly mislabeled and another 10 (2.5\%) were ambiguous. This suggests a conservative uncertainty estimate of 1–2\%, with a possible upper bound of ~4\% if ambiguous cases are included.
To complement the manual boundary check, we performed a robustness test repeating the UMAP + HDBSCAN clustering process with different random seeds (see supplementary material, Table S1). The observed label fluctuations were
small but slightly higher than those observed by manual inspection.
Because these variations may reflect clustering sensitivity rather than true ambiguity, we conservatively adopt the manually derived uncertainty estimate
(1–2\% around the cluster boundaries) as the benchmark for model fidelity.

\paragraph{Forward Feature Selection.}\label{sec:ffs}
Using different classifiers (namely logistic regression and gradient acceleration) and five-fold stratified cross-validation, we added descriptors one at a time until the test set AUC equaled the noise floor established above. Performance saturated after three characteristics,\(q_6\), entropy (\(\bar S_i\)), and \(p_2\)—reaching 0.9993, and plateaus near 0.9994–0.9995 thereafter (dashed line in the supplementary material, Fig. S15).
This AUC level realistically corresponds to roughly 1-10 potential
misclassifications out of 9000 atoms (Assuming $\Delta$ misclassifications are in the middle of the opposite class, and we have 50\% prior, $\mathrm{AUC} \approx 1 - (\Delta \times 2250)/(4500\times4500)$, which gives $\Delta \sim 10^{0..1}$). This is
consistent with our manual estimate of uncertainty based on inspection of the UMAP cluster boundary. Therefore, these three descriptors are adopted for all subsequent model explanations, and including more than three features risks overfitting to label noise rather than capturing additional meaningful structure.

\paragraph{Conditional-Independence Screening}\label{ci_screen}
To cross-validate the result of forward feature selection and to test whether any descriptor adds \emph{unique} information once all others are known, we applied a Gaussian Fisher-$Z$ partial correlation test between each feature and the binary \texttt{Cluster\_Label} with conditioning on the remaining descriptors. 
After correction for false discovery rate ($\alpha = 0.05$),
only \(q_6\), \(\bar S_i\), and \(p_2\) remained conditionally dependent throughout all temporal regimes; see supplementary material, Fig. S16.
Some of the other OPs, such as enthalpy or $q_8$, appeared relevant at specific times but did not maintain conditional dependence to the \texttt{Cluster\_Label}. Descriptors such as Voronoi volume and face count were
rendered redundant once the core predictive variables were included.

\begin{figure}[htbp]
    \centering

    \begin{subfigure}[t]{0.32\textwidth}
        \includegraphics[width=\linewidth]{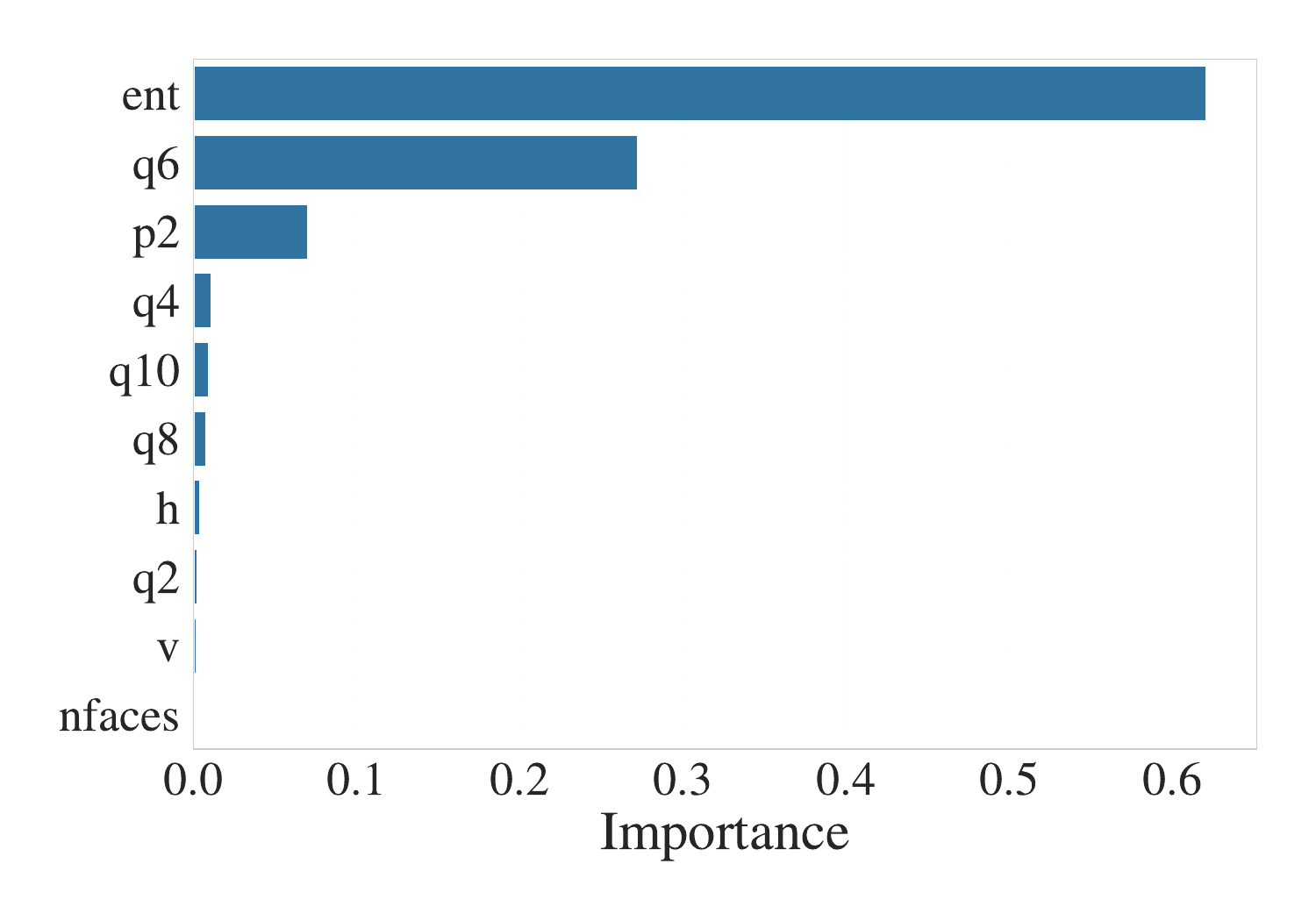}
        \caption*{\textbf{(a)}}
    \end{subfigure}
    \hfill
    \begin{subfigure}[t]{0.32\textwidth}
        \includegraphics[width=\linewidth]{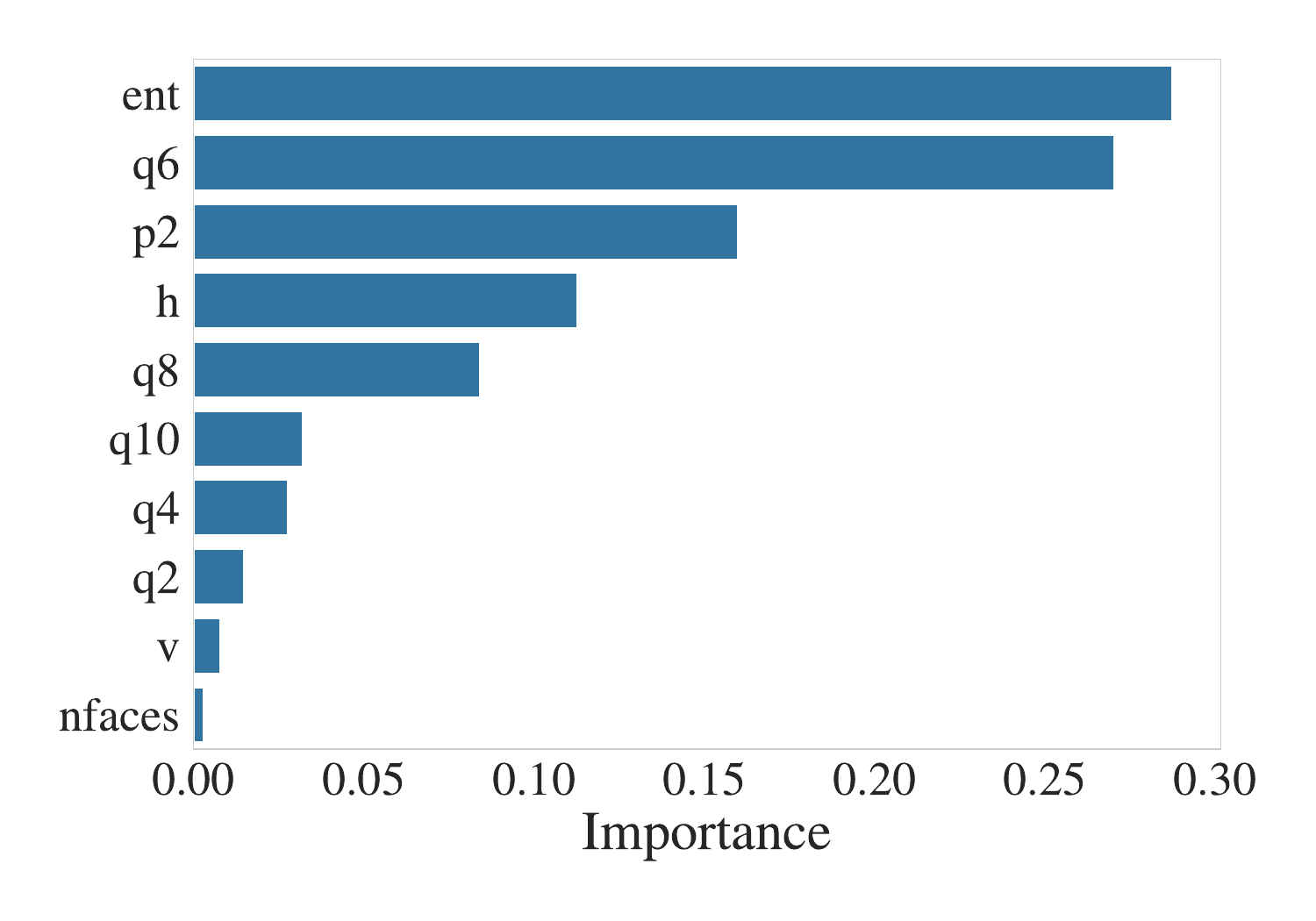}
        \caption*{\textbf{(b)}}
    \end{subfigure}
    \hfill
    \begin{subfigure}[t]{0.32\textwidth}
        \includegraphics[width=\linewidth]{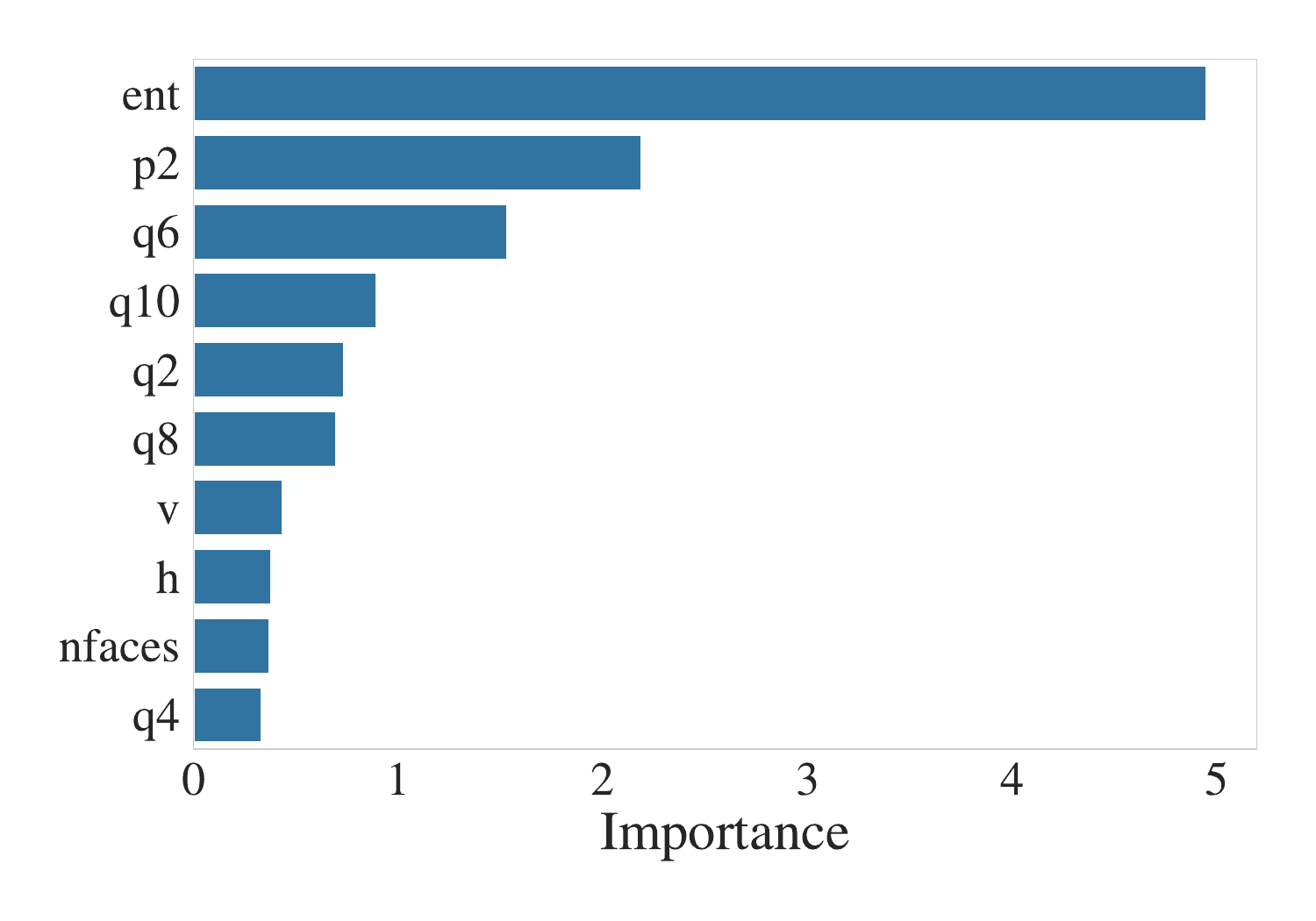}
        \caption*{\textbf{(c)}}
    \end{subfigure}

    \caption{Feature importance for (a) Gradient Boosting, (b) Random Forest, and (c) Logistic Regression. Entropy (\texttt{ent}), $q_6$, and $p_2$ consistently emerge as the most important features.}
    \label{fig:var}
\end{figure}

\paragraph{Feature Importance and Interpretability}
To further assess which features drive the classification of amorphous versus crystalline environments, we computed feature importance scores using three supervised models trained on the same timestep used in the forward feature selection. Figure~\ref{fig:var} presents the results for (a) logistic regression, (b) Random forest, and (c) Gradient boost. In all models, entropy (\texttt{ent}), $q_6$, and $p_2$ consistently rank among the most discriminant classification measures (Fig. \ref{fig:var}). This reinforces the outcome of forward feature selection and CI screening, which both identified this compact set as sufficient and non-redundant. These features, grounded in thermodynamics and symmetry, capture the key physical distinctions between disordered and ordered environments at early stages of crystallization. In the next subsection, we analyze how their ability to discriminate between phases evolves throughout the simulation trajectory.

\paragraph{SHAP analysis.}\label{sec:shap}
To further quantify and interpret these findings, SHAP (SHapley Additive
Explanations) was applied to the Gradient Boosting classifier trained on
\texttt{Cluster\_Label}. Unlike forward feature selection or traditional global feature-importance metrics, SHAP offers atomic-level explanations by
quantifying exactly how each descriptor contributes to pushing the model's
prediction toward crystalline (label 1) or amorphous (label 0) states on a
per-atom basis. This granularity is especially valuable for complex polymeric systems, where the same OP can assume different roles depending on local atomic environments.

\begin{figure}[htbp]
  \centering

  \begin{subfigure}[t]{0.47\textwidth}
    \includegraphics[width=\linewidth]{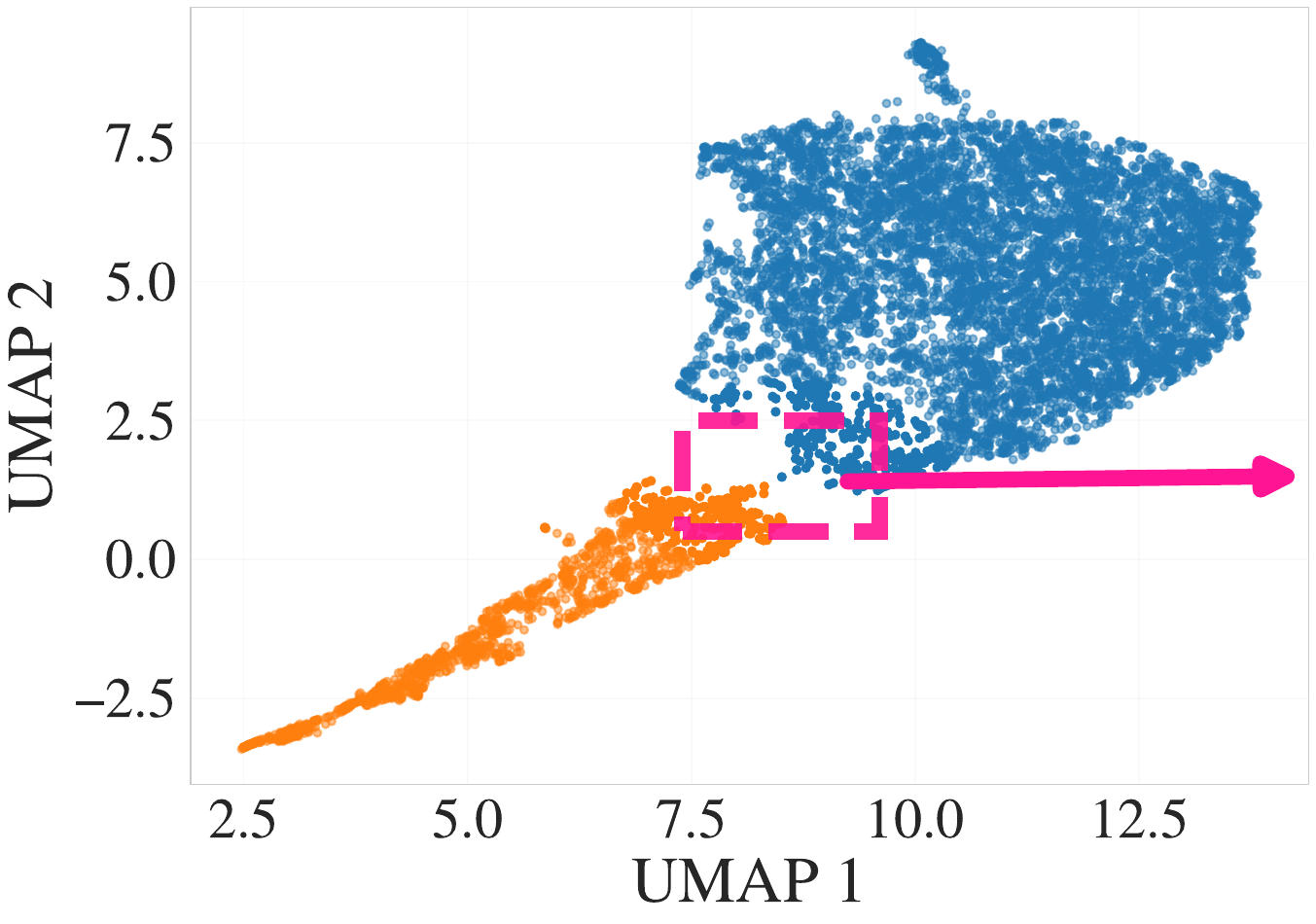}
    \caption*{\textbf{(a)}}
  \end{subfigure}
  \hfill
  \begin{subfigure}[t]{0.47\textwidth}
    \includegraphics[width=\linewidth]{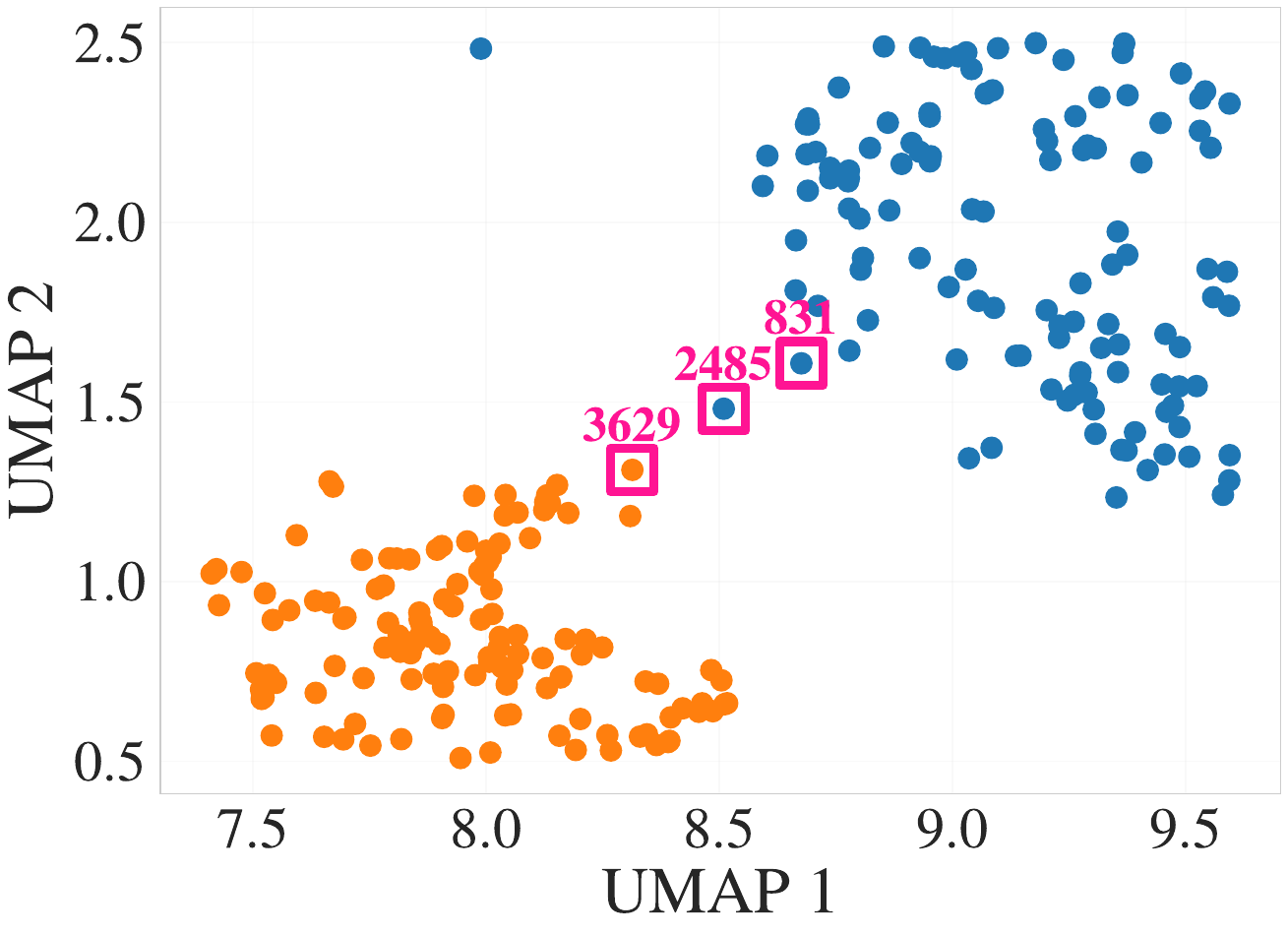}
    \caption*{\textbf{(b)}}
  \end{subfigure}

  \vspace{0.8em}

  \begin{subfigure}[t]{0.47\textwidth}
    \includegraphics[width=\linewidth]{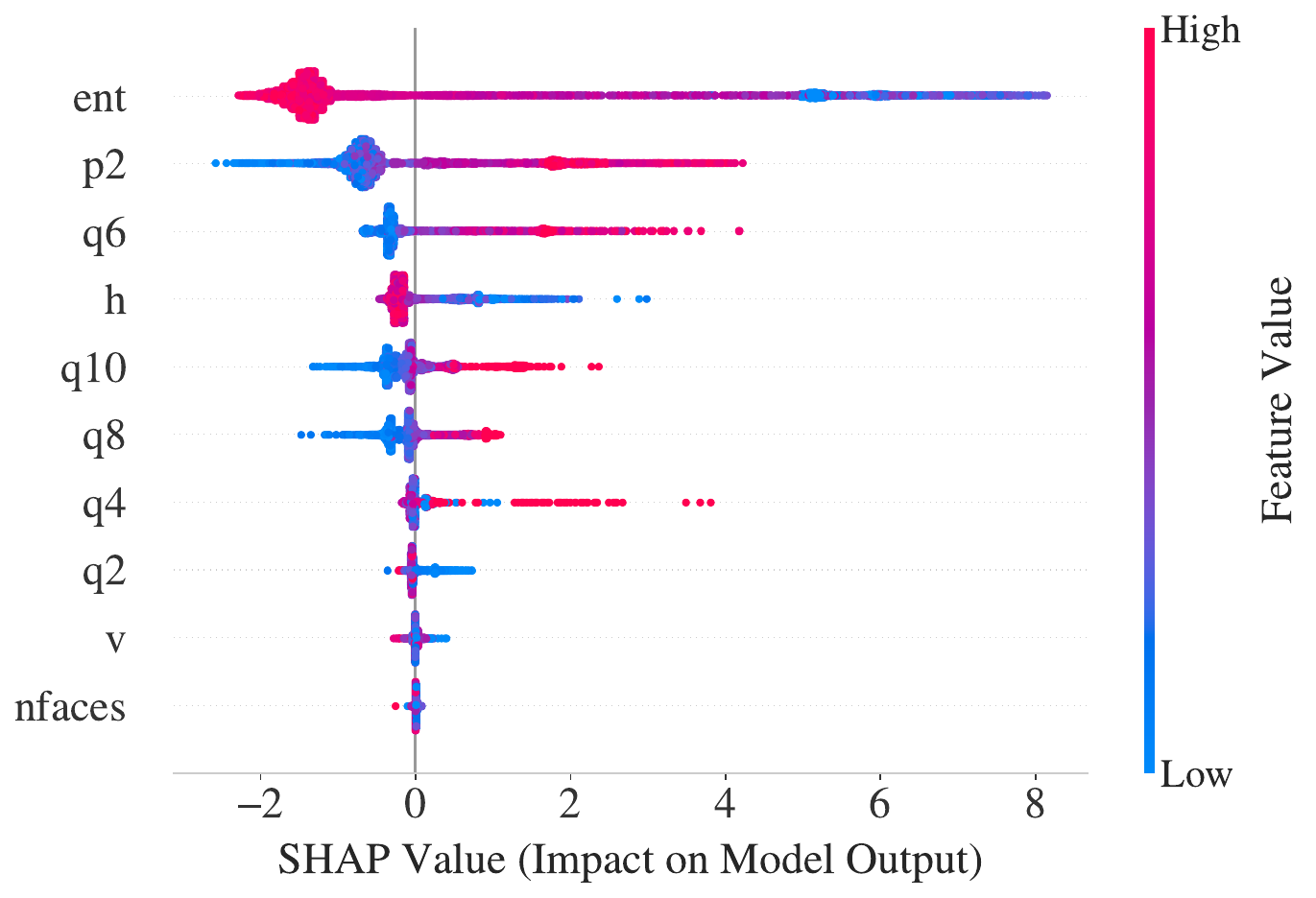}
    \caption*{\textbf{(c)}}
  \end{subfigure}
  \hfill
  \begin{subfigure}[t]{0.47\textwidth}
    \includegraphics[width=\linewidth]{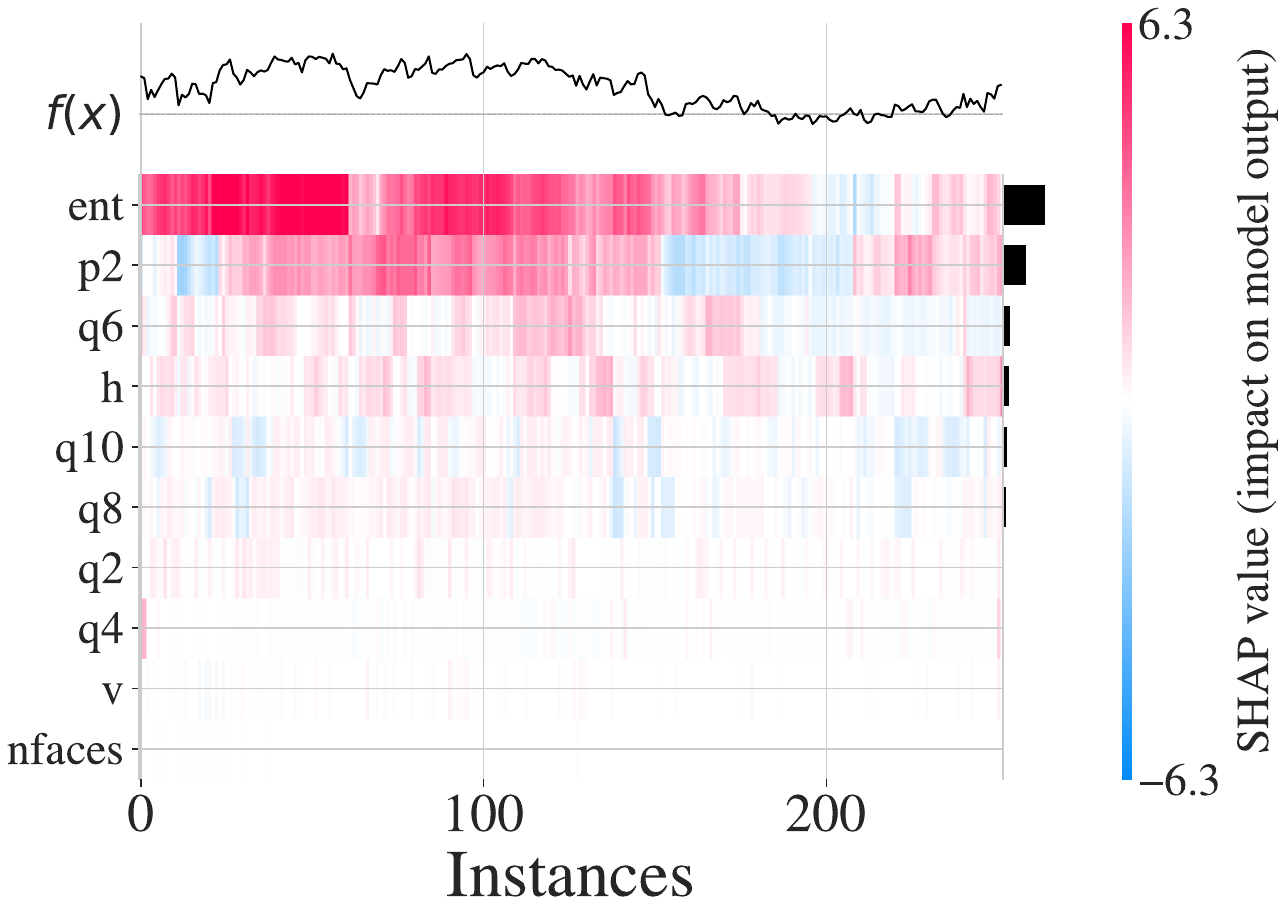}
    \caption*{\textbf{(d)}}
  \end{subfigure}

  \vspace{0.8em}

  \begin{subfigure}[t]{0.7\textwidth}
    \includegraphics[width=\linewidth]{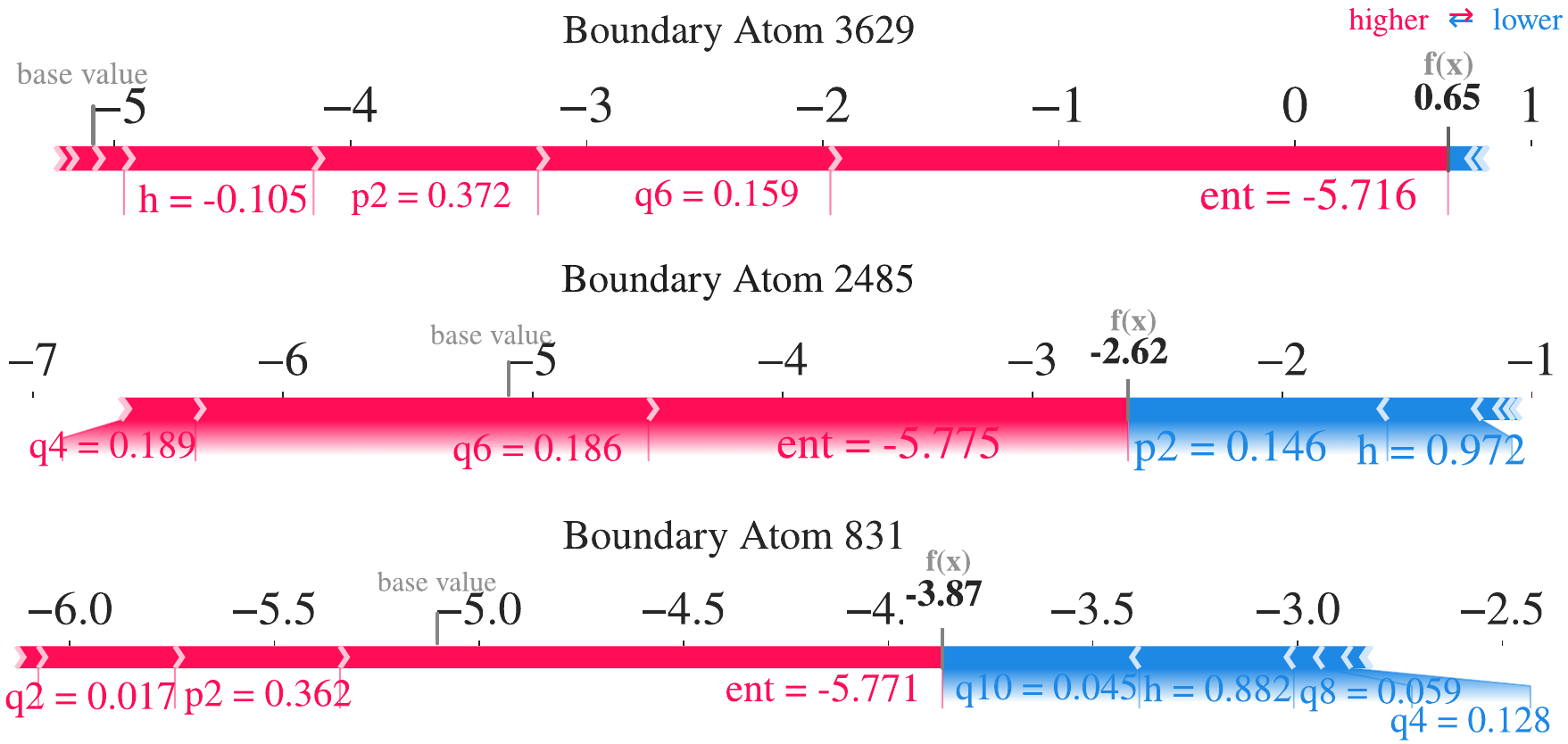}
    \caption*{\textbf{(e)}}
  \end{subfigure}

  \caption{
    SHAP-based interpretation of model predictions. 
    (a) UMAP with HDBSCAN labels; boundary atoms highlighted. 
    (b) Zoomed view showing three informative atoms. 
    (c) SHAP summary plot identifies entropy, $q_6$, and $p_2$ as key features. 
    (d) SHAP heatmap reveals local variation in feature importance. 
    (e) SHAP force plots for selected atoms show how entropy, symmetry, and enthalpy drive individual predictions, consistent with spatial context (see supplementary material, Fig.~S13).
  }
  \label{fig:shap}
\end{figure}

For example, entropy (\texttt{ent}) strongly drives predictions toward
crystallinity within highly ordered domains, as seen clearly in
Figure~\ref{fig:shap}(c), the global SHAP summary plot. Here, entropy
consistently emerges as the dominant contributor, followed closely by
metrics based on symmetry such as $q_6$ and $p_2$. However, near the
amorphous–crystalline boundary (Figure~\ref{fig:shap}(b,d)), entropy's
influence becomes context-dependent and is frequently modulated by other
characteristics such as enthalpy ($h$) and BOO parameters.

To explicitly illustrate this local heterogeneity, Figure~\ref{fig:shap}(d)
shows a SHAP heatmap for atoms within the zoomed-in boundary region highlighted in panel (b), bright red regions (high positive SHAP values) indicate features strongly reinforcing crystallinity, especially entropy, and, to a lesser extent, $p_2$ and $q_6$. In contrast, intermittent blue streaks reflect noncrystalline signals from high-enthalpy values and weaker symmetry cues, revealing microstructural nuances near the crystal boundaries invisible to global analyses.

The most detailed insights emerge from the single-atom SHAP force plots
(Figure~\ref{fig:shap}(e)). For boundary atom 831, despite strongly crystalline signals from entropy and moderate $p_2$, the combined effect of irregular enthalpy ($h = 0.88$) and lower orientational OP of the bond ($q_4, q_8, q_{10}$) pushes the prediction of the model toward the amorphous state, illustrating the subtle interaction of conflicting structural signals at cluster boundaries. Atom 2485 further highlights how learning from the clustering label can inform the
perceived importance of specific OPs: while entropy and $q_6$ consistently support crystalline classification, surface-sensitive descriptors ($h$ and $p_2$) ultimately dominate, resulting in an amorphous particle. Lastly, the 3629 boundary atom exemplifies a scenario in which entropy, $q_6$, $q_4$, and $p_2$ coherently align, decisively outweighing minor geometric irregularities and affirming a crystalline classification with near certainty. These same three atoms were previously studied in their real-space configuration in supplementary material Fig.~S13, where they were shown to reside at the morphological interface between clusters, further strengthening their role as informative probes for the model's ability to interpret features at the boundaries of the crystalline and amorphos regions. 

In general, SHAP analysis also reinforces the three-variable underlying rationale:
The entropy, $q_6$, and $p_2$ collectively form a robust yet compact feature set
whose individual contributions dynamically adjust according to the local structural context. This adaptability ensures the sensitive detection of heterogeneous nucleation behaviors across various atomic environments. It has the potential to enhance interpretation and predictive accuracy in the study of polymer crystallization.

\subsection{Introduction of the crystallinity index ($C$-index)}
\label{subsec:cryst_index}

\paragraph{Minimal descriptor set.}
Because \texttt{Cluster\_Label} itself is subject to labeling noise
($\approx$2–3 \% misassignments from manual 3D inspection), we asked whether a \emph{smaller} subset of descriptors could reproduce labels within this tolerance.  
Using only $\{q_6, p_2, \bar S_i\}$ (examining also a few next-ranked CI
survivors, mostly $\{h, q_8\}$), a gradient boost classifier still reaches an AUC of 0.979 (within the intrinsic labeling error) and an adjusted Rand index of 0.93 versus the full model.
This shows that the three variables selected by CI are sufficient for
practical classification, which motivates their use in the $C$-index introduced
in the following.

Building on the CI screen and minimal-descriptor test, we define a continuous \emph{Crystallinity Index} ($C$-index) from the three most informative and mutually non-redundant predictors, $q_6$, $p_2$, and \mbox{$\bar S_i$}.
Although the CI analysis identified $h$ and $q_8$ as secondary contributors, the
exploratory models showed that they add $<0.3 \%$ to AUC once $q_6$, $p_2$, and $\bar S_i$ are present; therefore, we keep the $C$-index parsimonious.

\paragraph{Definition of the $C$-index.} 
A logistic regression model was trained using these features to predict the
binary \texttt{Cluster\_Label}, and the resulting probability of an atom being
classified as crystalline (label 1 in \texttt{Cluster\_Label}) is taken as its
$C$-index.  This straightforward probabilistic model yields a scalar value
ranging from 0 (clearly amorphous) to 1 (clearly crystalline) and provides a
physically interpretable local order. Most importantly, the $C$-index
introduces a clean separation of phases that is difficult to achieve with
individual descriptors alone, especially in the early and transitional stages
of crystallization.

\begin{figure}[htbp]
    \centering

    \begin{subfigure}[t]{0.32\textwidth}
        \includegraphics[width=\linewidth]{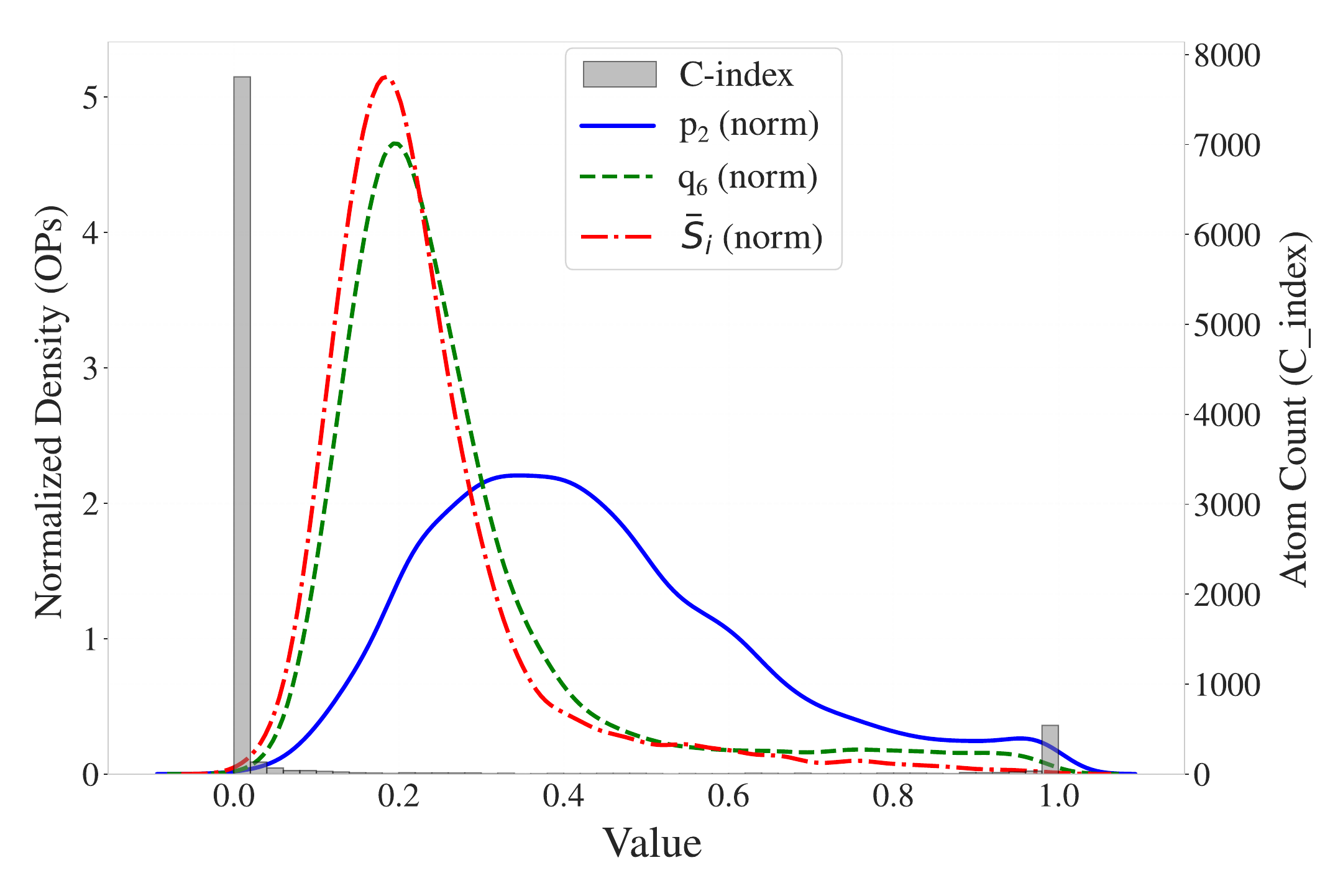}
        \caption*{\textbf{(a)}}
    \end{subfigure}
    \hfill
    \begin{subfigure}[t]{0.32\textwidth}
        \includegraphics[width=\linewidth]{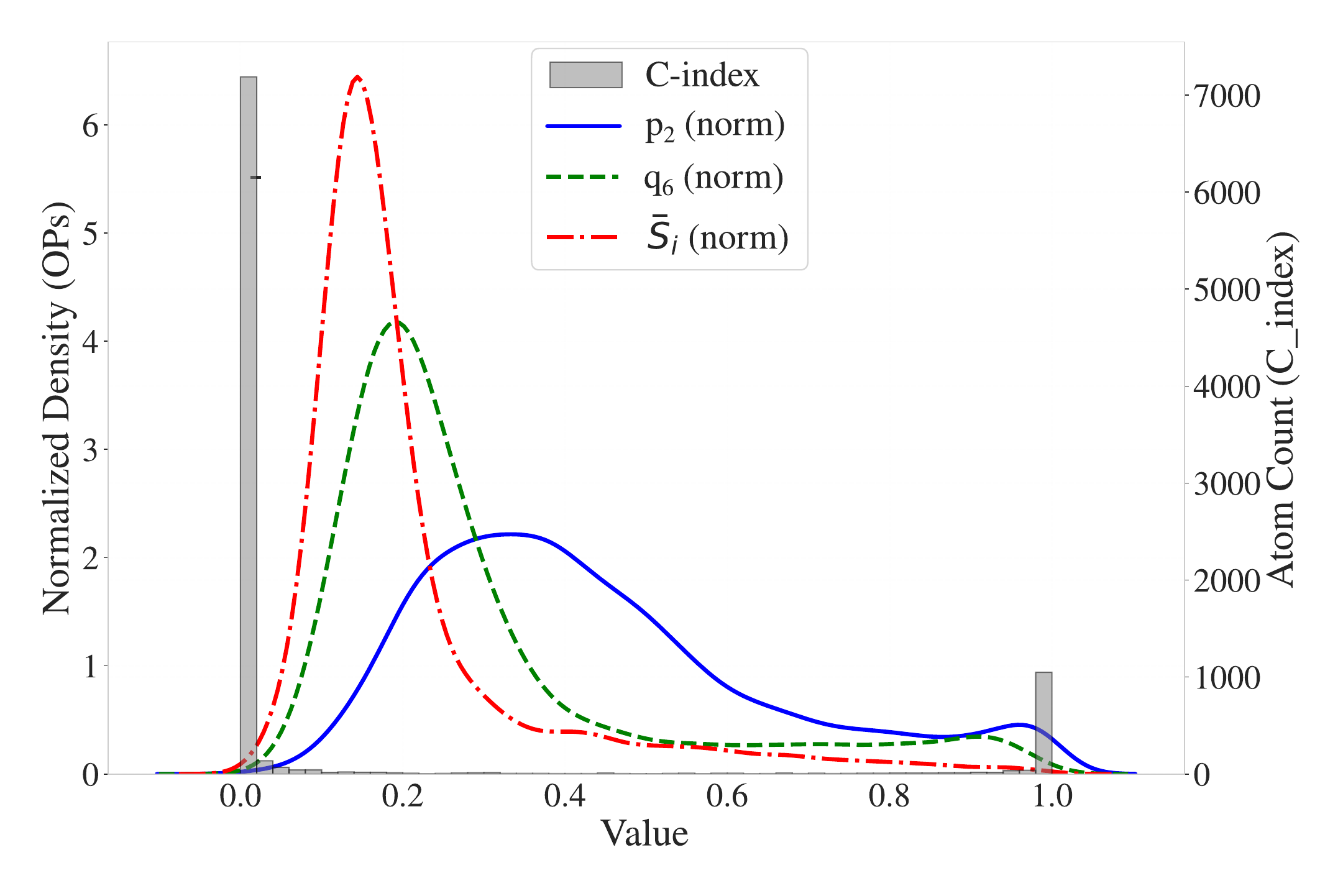}
        \caption*{\textbf{(b)}}
    \end{subfigure}
    \hfill
    \begin{subfigure}[t]{0.32\textwidth}
        \includegraphics[width=\linewidth]{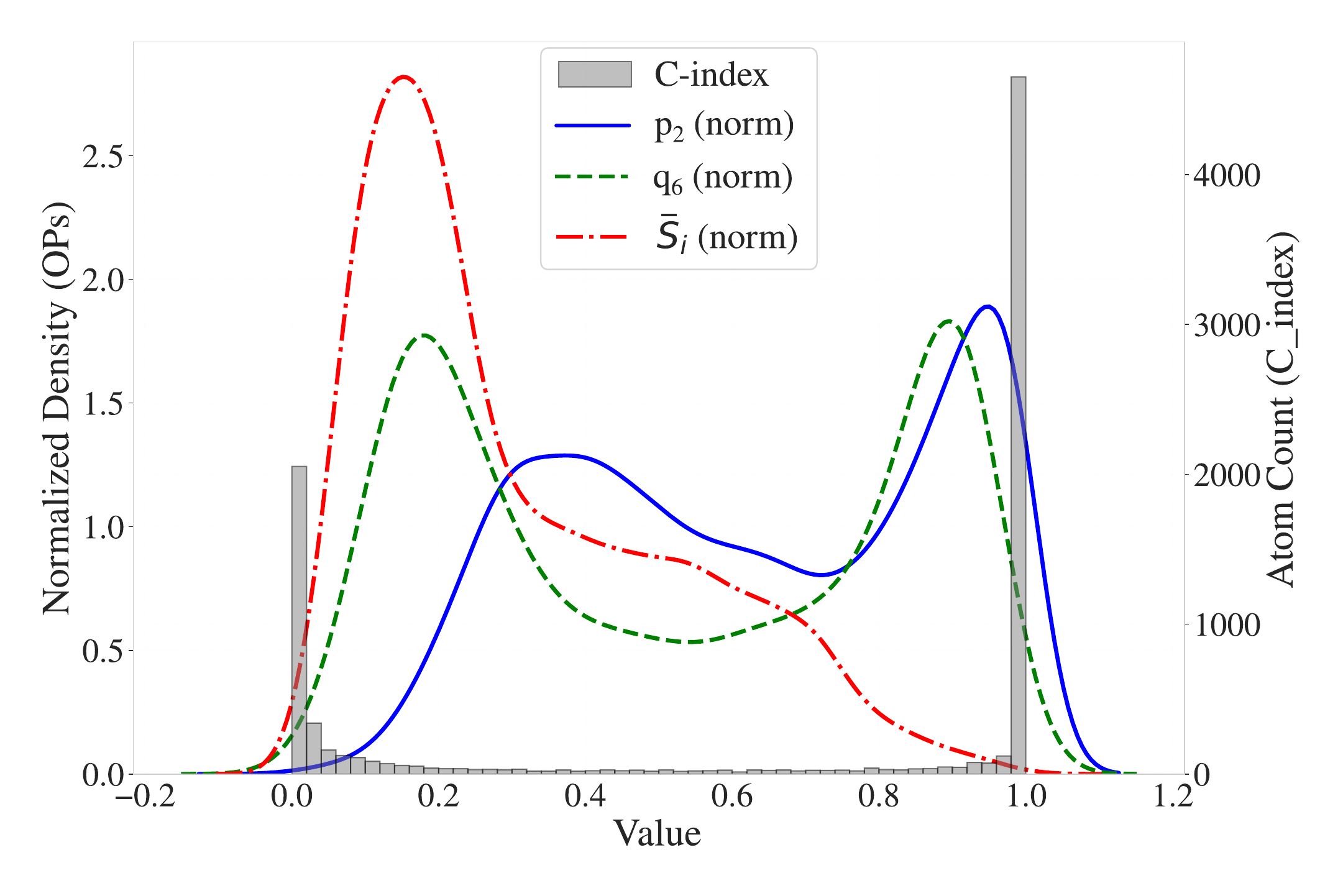}
        \caption*{\textbf{(c)}}
    \end{subfigure}

    \caption{Distributions for $C$-index and the most important OPs ($q_6$, $\bar S_i$, $p_2$) at different time steps. 
    (a) Transition onset ($t_{tr}$), (b) mid-stage growth ($t_{mid}$), and (c) steady-state crystallization ($t_{ss}$). Gray bars represent the histogram of the $C$-index (right y-axis), while colored curves show KDE-normalized distributions of $p_2$, $q_6$, and entropy (left y-axis). The $C$-index exhibits sharper phase distinction than any individual descriptor, avoiding the need for manual thresholding, especially in early and transitional regimes.}
    \label{fig:cindex_kde_final}
\end{figure}

Figure~\ref{fig:cindex_kde_final} illustrates the distribution of $C$ index values and normalized densities of the three input features in three representative time steps: the onset of nucleation ($t_{tr}$), the mid-growth phase ($t_{mid}$) and steady-state crystallization ($t_{ss}$). As shown in panel (a), during early nucleation, the distributions of $q_6$, $p_2$ and
entropy overlap significantly and lack clear thresholds. In contrast, the $C$-index shows a sharp unimodal distribution near zero, with a small but distinct shoulder that indicates the emergence of crystalline atoms. At $t_{mid}$ (panel b), the $C$-index becomes bimodal, clearly separating the amorphous and crystalline populations, unlike the individual order parameters.
The system exhibits well-developed crystal growth by $t_{ss}$ (panel c), and the $C$-index strongly peaks near 1 for crystalline atoms. These results emphasize the robustness and interpretability of the $C$-index compared to individual descriptors, particularly in noisy or intermediate regimes.
The reliability of the $C$-index in all stages of crystallization is supported by AUC scores that exceed 98\% at all time steps, with the table given in the supplementary material, Table ~S2.

\begin{figure}[ht]
  \centering
  \includegraphics[width=\textwidth]{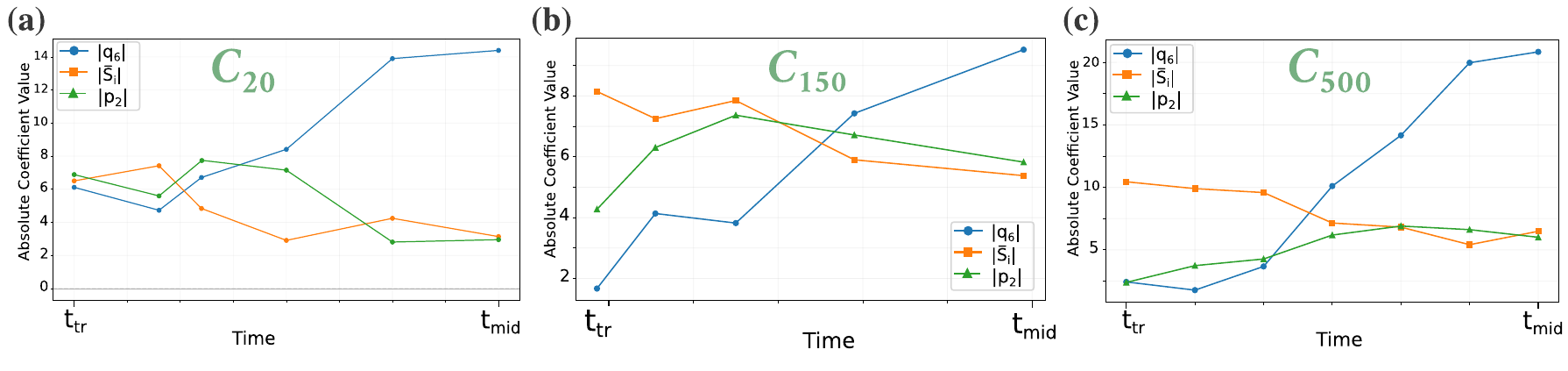}
  \caption{
    Evolution of absolute logistic regression coefficients for $q_6$ (blue),
    entropy $\bar S_i$ (orange), and $p_2$ (green) over characteristic
    time steps between ($t_{tr}$, to $t_{mid}$) for chain lengths C${20}$
    (left), C${150}$ (center), and C${500}$ (right). Early on ($t_{tr}$),
    entropy dominates; through the time, $q_6$ and $p_2$ increase in
    importance; at mid-grown step ($t_{mid}$), $q_6$ becomes the primary
    discriminator of crystallinity.
  }
  \label{fig:coeff_evolution}
\end{figure}

To better understand the temporal robustness of the $C$ index, one can examine how the logistic regression coefficients for $p_2$, entropy ($\bar S_i$) and $q_6$ evolve over time and vary with chain length.
Figure~\ref{fig:coeff_evolution} shows the absolute values of these
coefficients in three characteristic stages: nucleation onset ($t_{tr}$),
mid-growth phase ($t_{mid}$) and steady-state crystallization ($t_{ss}$) for
three different chain lengths: C${20}$, C${150}$, and C${500}$.
Several important trends emerge.
At early nucleation ($t_{tr}$), entropy has the largest coefficient in the longer entangled chains (C${150}$ and C${500}$), indicating that low-entropy fluctuations are the most discriminative signal of nascent crystallinity.
Interestingly, the difference between the entropy and the other coefficients is more pronounced in C${500}$ compared to C${150}$, suggesting a stronger entropic signal in highly entangled systems.
In contrast, for unentangled C$_{20}$ chains, the three characteristics
contribute nearly equally, confirming reduced entropic frustration in the
nucleation process.

During the growth phase between $t_{tr}$ and $t_{mid}$, a change occurs: the
weight of entropy begins to decline while the contributions from $q_6$ and
$p_2$ increase, particularly in C${150}$ and C${500}$. For C$_{20}$, both the entropy and $p_2$ decrease slightly while $q_6$ continuously grows. This
reflects the physical evolution of clusters; that is, the increase in the
orientation order of the bonds ($q_6$) as the local structure and symmetries
develop.

By the time the system reaches $t_{mid}$, $q_6$ uniformly dominates across all lengths of the chain. This reflects the physical evolution of the crystalline phase: as the clusters mature, they begin to exhibit strong internal symmetry, which is precisely what $q_6$ quantifies through the orientation of the bond.
Although entropy and $p_2$ remain meaningful descriptors, crystalline regions still maintain low entropy and high chain alignment; increasing structural regularity within clusters makes $q_6$ the most discriminative characteristic at this stage. In essence, $q_6$ captures the developing symmetry of the crystalline lattice, whereas entropy and $p_2$ describe aspects that have already plateaued. 
This behavior comes from the fact that the logistic regression model learns to exploit statistical differences in the feature space between the two classes.
Therefore, the model naturally assigns a higher weight to $q_6$ because it
continues to offer the strongest contrast between amorphous and crystalline
atoms as crystallization progresses.

Despite these shifts in the weights per characteristic, a single $C$ index
model trained at one time step (e.g. $t_{mid}$) still achieves excellent
classification performance when applied to other time steps
(shown by the AUC table in the supplementary material, Table ~S2).
In other words, the decision boundary is defined by
\begin{equation}
  w_{p_2}^{(t_{\text{mid}})}\, p_2 +
  w_{\bar{S}_i}^{(t_{\text{mid}})}\, \bar{S}_i +
  w_{q_6}^{(t_{\text{mid}})}\, q_6 +
  b^{(t_{\text{mid}})}
  = 0~,
\end{equation}
continues to separate crystalline and amorphous atoms with high fidelity over time. This transferability arises because, although the optimal weights adjust to reflect which feature has the greatest variance between classes at each snapshot, the overall geometry of the two populations in $(p_2, \bar S_i, q_6)$ space remains qualitatively similar: crystallized atoms consistently occupy a region of low entropy, high $q_6$ and $p_2$, regardless of whether they are nascent clusters or fully formed crystals. As long as the hyperplane from $t_{mid}$ still ranks truly crystalline points above amorphous ones, the AUC of the ROC remains high.

In other words, the evolution of the per-timestep coefficient captures the
subtle change in which characteristic is the most 'informative' at that
instant, but does not imply that the separating plane rotates so drastically
that it cannot classify atoms correctly over time. Consequently, a single
well-trained $C$ index (based on a carefully selected subset of the feature
space) can be used to monitor nucleation and growth in real time, without
retraining in every time step. This robustness makes the $C$-index particularly suited for large-scale simulations where computational efficiency and the ability to interpret interfacial atoms are critical.

By construction, the $C$ index inherits physical interpretation, generalization,
and robustness. It offers a principal scalar summary of local crystallinity without requiring discrete thresholds or retraining at each timestep, making it well suited for real-time tracking of nucleation and growth in large‐scale simulations.

\section{Conclusion}
\label{sec:Conclusion}

A data-driven framework for identifying and tracking polymer crystallization in large-scale molecular dynamics simulations is presented and discussed. The approach begins by defining a comprehensive suite of conventional and thermodynamic-like order parameters, including local density from Voronoi tessellation, nematic alignment ($p_2$), entropy ($\bar{S}i$), enthalpy ($H_i$), and bond orientational order parameters ($q_{\ell}$).
These descriptors are combined into a high-dimensional feature vector. Using UMAP, we show that one can generate a low-dimensional representation that captures a clear separation between phases, and with HDBSCAN clustering, we generate high-quality labels to classify atoms as crystalline or amorphous. For model explanation, the generated high-quality labels were used in a supervised setting. $q_6$, $\bar{S}_i$, and $p_2$ were consistently selected as the most discriminant measures at different stages of crystallization or chain lengths.
After conditional independence tests and error analyses, we show that these three characteristics reproduced the UMAP plus clustering result \textit{almost perfectly} in a supervised setting. Using logistic regression probability, the crystallinity index, the $C$-index, is introduced as a continuous, interpretable scalar ranging from 0 (amorphous) to 1 (crystalline).

In addition, we demonstrate that the index $C$, trained only using a few or a few snapshots in the middle time interval, sharply distinguishes phases even in the early and intermediate nucleation and crystallization regimes, achieving AUCs $> 0.98$ at all stages, while individual OP distributions of the early and late stages show a large overlap between phases.
Analyzing the coefficient evolution in time, by retraining the $C$-index model,
entropy dominates at nucleation onset in entangled chains, while in short unentangled chains (e.g., C$_{20}$), the three descriptors contribute more evenly. As crystallization progresses, $p_2$ takes over during growth, and bond orientational symmetry prevails at steady state across all chain lengths.

In summary, the introduced workflow successfully captures a high-dimensional OP space and
provides a computationally efficient strategy for monitoring crystal nucleation and growth.
The $C$-index, in particular, offers a reusable threshold-free measure of local order that can be applied without retraining at every time step, making it ideal for large‐scale or on‐the‐fly analyses. We anticipate that the methodologies and framework introduced herein can be used or extended to other problems where many noisy OPs capture different aspects of a needed abstraction, irrespective of material or simulation protocols involved.

\section*{Supplementary Material}

Additional details are provided in the supplementary material, including pairwise descriptor correlations, additional insights into OPs, UMAP and clustering hyperparameter tuning results, comparisons across dimensionality reduction techniques, and a comprehensive analysis of supervised model training, feature selection, and target label consistency. Supporting figures and tables are included for each section.

\section*{Acknowledgments}
The authors acknowledge the Texas Advanced Computing Center (TACC) at the University of Texas at Austin for providing high-performance computing resources used in the large-scale molecular dynamics simulations. Post-processing and data analysis were performed using computational infrastructure provided by the ISAAC at the University of Tennessee, Knoxville. Financial support was provided by the Materials Research and Innovation Laboratory (MRAIL) at the University of Tennessee, Knoxville.


\section*{Conflict of Interest}
The authors have no conflict of interest to disclose.


\section*{Data Availability}
Data supporting the findings of this study are available from the corresponding author on a reasonable request.

\bibliographystyle{unsrt}
\bibliography{references}

\end{document}